\documentclass[prd,
a4paper,
nofootinbib,
reprint,
]{revtex4}
\usepackage{graphicx}
\usepackage{amsfonts}
\usepackage{amssymb}
\usepackage{amsmath}
\usepackage{slashed} 
\usepackage{enumerate}
\usepackage{color}
\usepackage{hyperref}
\def\eq#1{{eq.~(\ref{#1})}}

\definecolor{oucrimsonred}{rgb}{0.6, 0.0, 0.0}
\definecolor{persianblue}{rgb}{0.11, 0.22, 0.73}
\definecolor{forestgreen}{rgb}{0.13,0.35,0.13}
 \hypersetup{colorlinks, citecolor=oucrimsonred, linkcolor=persianblue, urlcolor=oucrimsonred}
\def\hhref#1{\href{http://arxiv.org/abs/#1}{#1}} 
\newcommand{\be}{\begin{equation}}
\newcommand{\ee}{\end{equation}}
\newcommand{\bea}{\begin{eqnarray}}
\newcommand{\eea}{\end{eqnarray}}

\definecolor{oucrimsonred}{rgb}{0.6, 0.0, 0.0}
\begin{document}
\title[]{Charged neutron stars  and observational tests \\
 of a dark force weaker than gravity
}
\date{\today}
\author{M.\ Fabbrichesi
}
\author{A.\ Urbano 
}

\affiliation{INFN, Sezione di Trieste, Via  Valerio 2, 34127 Trieste, Italy }

\begin{abstract}
\noindent  We discuss the  possibility of  exploring an  unbroken $U(1)$  gauge interaction in the dark sector  by means of gravitational waves. Dark sector states charged under the dark force can give a macroscopic charge to astronomical bodies. Yet the requirement of having gravitationally bounded stars limits this charge to negligible values if the force has a sizeable strength. Gravitational tests are only possible if the dark force is weaker than gravity.
By solving the Einstein-Maxwell field equations, we study in detail an explicit model for dark charge generation and separation in a  neutron star.
Charged states originate from the decay of  neutrons inside the star into three dark fermions; we show that  in this model the equation of state  is consistent with limits on neutron star masses and tidal deformability.
We find that while the dark force can be observed in binary mergers (making them an   optimal observational test even though with limited precision), it is Debye screened in binary pulsars (for which more precise data exist).
The emitted radiation in the inspiral phase of a binary system is modified and the dark force tested at the  level of the uncertainty of the experimental detection. 
The test covers   a region where current limits on deviations from Newton inverse-squared law  come from geophysical and laser-ranging observations. 
 \end{abstract}
\maketitle
 
 \section{Motivations}
 
The evidence in favor of the existence of dark matter lends itself to a generalization in which such matter is but one member of an extended  dark sector  comprising many different states (see~\cite{dark_sector} for two recent reviews). Allowing interactions within this dark sector seems natural enough: the simplest one to be a $U(1)$ gauge interaction, modelled as a dark analog of electromagnetism, under which all states in the dark sector are charged~\cite{Ackerman:mha}. 

Various constraints apply to the dark matter component of the dark sector, the most relevant in our case being that its mass must be rather large if the dark force has an appreciable strength. This comes about in order to preserve the essentially collisionless nature of dark matter  in the presence of the dark (long-range) interaction~\cite{Ackerman:mha}. The value of the relic density provides the other  essential ingredient in the determination of the dark sector parameters~\cite{Ackerman:mha,Barducci:2018rlx}.
Other constraints are model dependent and mainly about the states providing a link between the dark and the ordinary sectors~\cite{Hoffmann:1987et,Biswas:2016jsh}.

Is it possible to probe experimentally this hypothetical dark force? Currently, there are no direct limits because only ordinary matter enters the experimental constraints and, if the symmetry is unbroken, the dark force is necessarily contained only in higher-order operators (of dimension equal to or larger of six) in which the dark force coupling is modulated by the effective scale of the operator and the details of the portal between the dark and the standard-model sectors.

The study of gravitationally bounded states like a binary system  of two neutron stars---made possible by the detection of gravitation waves~\cite{TheLIGOScientific:2017qsa}---opens new possibilities that could provide such a probe if the stars contain a sufficient number of states carrying dark charges. This has been discussed in~\cite{Sagunski:2017nzb,Kopp:2018jom}---whose results  we reproduce as we pursue our line of reasoning (see also \cite{Alexander:2018qzg,Choi:2018axi,Hook:2017psm,Huang:2018pbu}). We consider an explicit (albeit minimal)  model of the dark sector~\cite{Barducci:2018rlx} to study in detail how the dark charged states can be produced, how their presence modifies the equation of state  and how their charged are separated in the case of neutron stars. They affect   the quadrupole radiation and can generate a dipole term in the radiation emitted by the binary system.

As in the electromagnetic case, the requirement of having gravitationally bounded stars limits the dark charge carried by them to negligible values when the force has a sizeable strength~\cite{bally}.  The study of macroscopic astrophysical objects of mass $M$, carrying a dark charge $\mathcal{Q}$, can only help if the dark force coupling strength $\alpha$ is of the order of the effective gravitation interaction, that is, when
$\alpha \mathcal{Q}^2/G_N M^2$
is order 1 ($G_N$ is Newton constant). This condition restricts  the study of gravitational waves emitted by astronomical bodies to   dark forces that are  weaker than gravity. 

We therefore need to adjust our original motivation  because gravitational waves cannot  help in the range of strengths of the dark force we might be interested in to  alter significantly the properties of   the dark sector. The dark sector can  be studied by means of gravitation waves only if its self-interaction (in addition to gravity)  has a strength at most comparable to gravity. This super-weak dark force  can be excluded with increasingly higher  probability as the experimental uncertainties in gravitational-wave detection are narrowed down. Gravitational-wave tests cover  a region where current limits on deviations from Newton  inverse-squared law (see, for example, \cite{Adelberger:2003zx}) only come from geophysical and laser ranging observations which cannot be sensitive to dark sector interactions.

\section{Charged stars and other gravitationally bound system}\label{sec:DarkStars}

A star is  a sphere of gas held together by its own gravity. 
The gas pressure balances the force of gravity, and creates a situation of hydrostatic equilibrium.
Neutron stars are the most compact stars known in the Universe,
and their interior is almost entirely composed of neutrons.
Neutron stars form from the explosive death of massive stars (typically main-sequence stars with mass 
$M\gtrsim 8\,M_{\odot}$), 
after their central region---no longer supported by the energy released in nuclear fusion processes---collapses under gravity causing protons and electrons to combine into neutrons. 
After this extreme condition occurs, the crushing force of gravity is balanced by the pressure of the neutrons. 

 We  are interested in the possibility that a neutron star carries a non-zero $U(1)$ dark charge. In such a situation, the condition of hydrostatic equilibrium changes with respect to the pure gravitational case.  We  review   in the next section
the hydrostatic equilibrium for the  well-known case of a generic charged star. This example serves as a prototype for the
 case (we are most interested in) of neutron stars---which we discuss  in section~\ref{sec:ChargedNS}, where the same equations are solved in the presence of  an explicit mechanism for making the star charged.

\subsection{Equilibrium conditions in Newtonian gravity}\label{sec:DarkStarsEq}

In Newtonian gravity, the hydrostatic equilibrium is described---assuming spherical symmetry---by the differential equation
 \begin{equation}\label{eq:NewEq}
\frac{dP(r)}{dr} = -\frac{G_Nm(r)\rho(r)}{r^2}~,
\end{equation}
 describing the balance between the net outward pressure force and the inward gravitational force on the infinitesimal fluid element at distance $r$. $P(r)$ and $\rho(r)$ are, respectively, the pressure and the mass density at position $r$ while $m(r)$ is the mass interior to the radius $r$.
Throughout this paper, we assume spherical symmetry. In eq.~(\ref{eq:NewEq}), the gradient pressure is negative. This means that, starting from a given value at the center of the star, the pressure decreases going outward, and it vanishes at the surface of the star. 

Since we are interested in charged stars, we assume a non-zero charge density $\rho_e(r)$. 
Eq.~(\ref{eq:NewEq}) gets modified by the presence of the charge in two ways. We find
 \begin{equation}\label{eq:NewEqCharge}
\frac{dP(r)}{dr} = -\frac{G_N\rho(r)}{r^2}\bigg[
m(r)  \underbrace{-2\pi r^3E(r)^2}_{\rm electrostatic\,pressure}
\bigg] + \underbrace{\rho_e(r)E(r)}_{\rm Coulomb\,force}~,
\end{equation}
where the radial electric field $E(r)$ is given in terms of the charge $Q(r)$ within the radius $r$ by
\begin{equation}
E(r) = \frac{Q(r)}{4\pi r^2}~,~~~~ Q(r) =  
4\pi\int_0^{r}d\tilde{r}\tilde{r}^2\rho_e(\tilde{r})~.
\end{equation}
In eq.~(\ref{eq:NewEqCharge}) the first correction comes from the presence of the additional contribution of the electrostatic pressure 
 while the second term accounts for the Coulomb force acting on the infinitesimal charged fluid element. The  electrostatic pressure is always opposite to gravity because the Coulomb repulsion works against  gravitational attraction. 
By increasing the charge density, the 
electrostatic pressure  grows up to the point at which it overcomes
the gravitational attraction. The sign of the pressure gradient in eq.~(\ref{eq:NewEqCharge}) becomes positive and it is no longer possible to find an equilibrium solution.
\begin{figure}[!htb!]
\begin{center}
	\includegraphics[width=.5\textwidth]{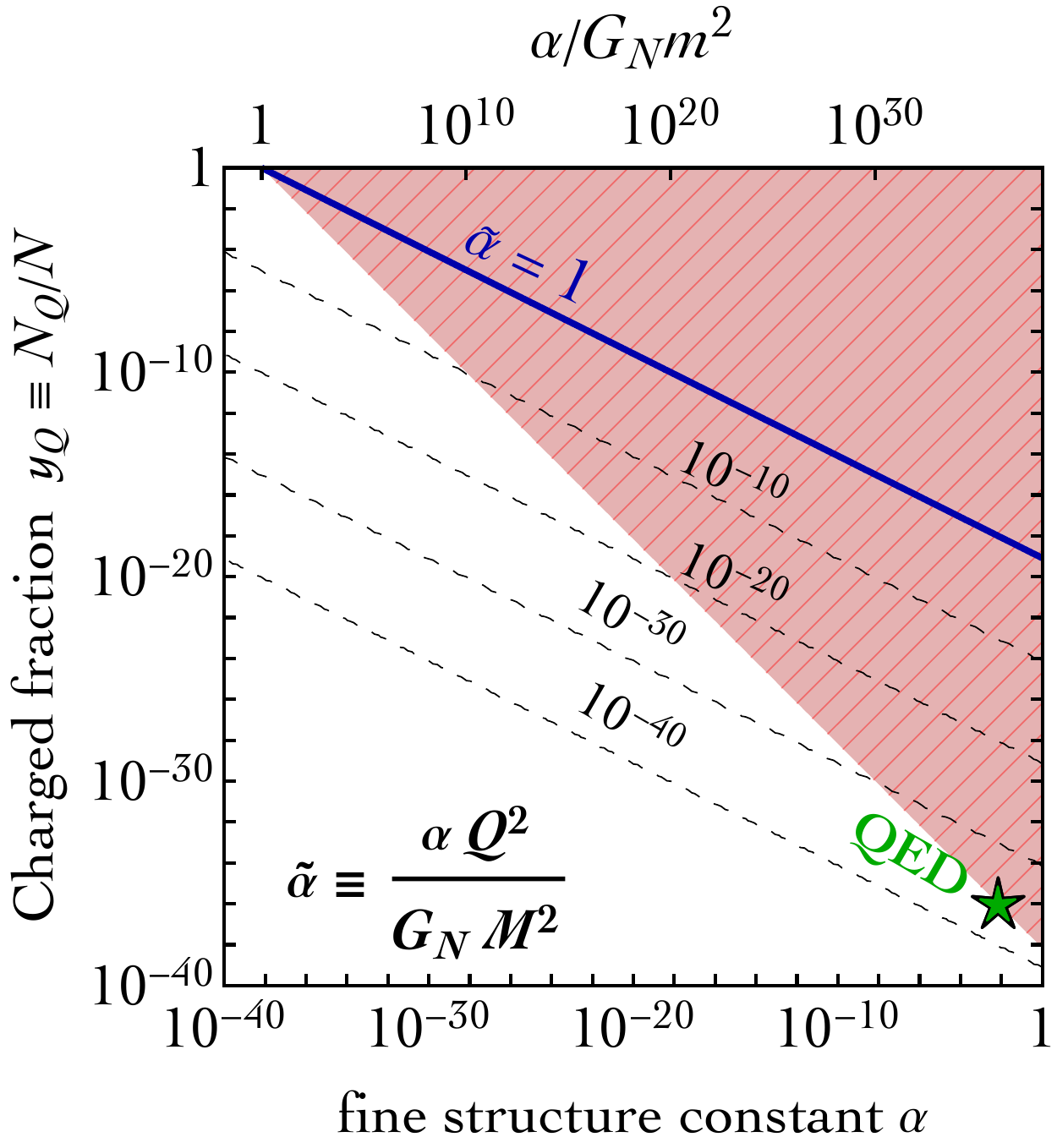}
	\caption{\em \label{fig:StarPlot} 
Structural limit on the maximal amount of charge that can be carried by a star in the plane $(\alpha, y_Q)$, where $\alpha$ is the fine structure constant of the $U(1)$ gauge interaction and $y_Q$ the fraction of the star that is charged.	
	In the red region, the electrostatic pressure is stronger then the pull of gravity, thus preventing  the
	formation of a charged compact object.  Diagonal dashed lines show contours of constant $\tilde{\alpha}$, that is the relevant parameter in the study of gravitational wave signals (see section~\ref{sec:GW}). The green dot marks the maximum fraction of a star that can be charged under conventional electromagnetic interactions ($\alpha = 1/137$).
 }
\end{center}
\end{figure}

To be specific, let us consider a star with mass $M$ and radius $R$. The star is constituted by $N$ particles with mass $m$, and $M = mN$. 
If we take $m = 1$ GeV, in order to have a solar-mass object we need $N \approx 10^{57}$. 
We assume that a fraction of the object is charged, and we define the total charge $\tilde{\mathcal{Q}} = e\mathcal{Q} = eN_Q \equiv e(y_Q N)$, where $e$ is the gauge coupling.\footnote{We use the notation of electromagnetism but we refer to a generic dark $U(1)$ gauge symmetry.} 
We also assume, for simplicity, constant mass-density distribution $\rho = M/(4\pi R^3/3)$, and we fix $R= 10$ km, which is the typical radius of a neutron star. 

In this simplified set-up, we only have two free parameters: the gauge coupling $e$ (equivalently, the fine structure constant $\alpha = e^2/4\pi$) and the charged fraction $y_Q$. We impose $dP/dr <0$ in eq.~(\ref{eq:NewEqCharge}). At distance $r=R$, it corresponds to the condition
 \begin{equation}\label{eq:BoundAlpha2}
 4\pi\alpha = e^2 < \frac{8G_N m^2 \pi R}{y_Q(y_Q G_N M + 2 R)}~.
  \end{equation}
The region shaded in red in fig.\,\ref{fig:StarPlot} does not satisfy the  relation in eq.~(\ref{eq:BoundAlpha2}). In this region, the Coulomb repulsion prevents  the formation of a charged star. The green star corresponds to the QED case with $\alpha = 1/137$. 
   We see that in this case the Coulomb repulsion is so strong that only a tiny fraction of the
   star can be charged, nanely $y_Q \simeq 10^{-36}$. 
   
   Since in QED the charge $e \simeq 1.6\times 10^{-19}$ C, we find the maximal total charge  
   $\tilde{\mathcal{Q}}_{\rm QED} \simeq 200$ C, in agreement with known results~\cite{bally,Ross}. This is indeed a very small charge---for comparison: the charge of a  AAA battery is about ten times larger.
   
   In general, it is possible to 
   have a sizeable fraction of the star that is charged only if the gauge coupling is super-weak. In fig.\,\ref{fig:StarPlot} we show contours of constant ratio 
   \begin{equation}\label{eq:Alpha}
   \tilde{\alpha} \equiv \frac{\alpha\mathcal{Q}^2}{G_N M^2} = \frac{\alpha y_Q^2}{G_N m^2}~.
   \end{equation}
   As we  discuss in section~\ref{sec:GW}, this parameter  could be tested with gravitational wave physics;
   it controls  the relative strength of the $U(1)$ dark gauge force     
   compared to gravity at the macroscopic level. The solid blue line corresponds to $\tilde{\alpha} = 1$, and $\tilde{\alpha} > 1$ indicates a dark force macroscopically stronger than gravity.  This region is entirely excluded by the structural bound discussed in this section.
   
  Fig.~\ref{fig:StarPlot}, therefore, conveys a trivial but important message: At the macroscopic level, gravity must dominate over other possible repulsive forces in order to ensure the existence of gravitational bound states. 
   
   The case of QED is, in this respect, emblematic. 
   Microscopically, 
   the electromagnetic repulsion between two particles with the same electric charge overwhelms the gravitational attraction (for all known charged particles with  mass $m$ we have $\alpha/G_N m^2 \ggg 1$) but at the macroscopic level (where, instead of $\alpha/G_N m^2$, what is relevant is  $\tilde{\alpha} = \alpha\mathcal{Q}^2/G_N M^2$) only values  $\tilde{\alpha}\lll 1$  are allowed. As it is clear from the top-left corner of fig.\,\ref{fig:StarPlot}, the only chance to obtain $\tilde{\alpha}\lesssim 1$ is to consider a super-weak coupling. In this case $\alpha \approx G_N m^2$ already at the microscopic level (the top axes in   fig.\,\ref{fig:StarPlot}), and one can push  $y_Q$ (and thus $\tilde{\alpha}$) to sizeable values.
    	
This argument, even if formulated at the Newtonian level,  shows that there exists a structural bound against the possibility of forming a charged star. 
The simple discussion proposed here---besides ignoring general relativistic corrections to Newtonian gravity, which are important for objects as compact as neutron stars---does not explain the microscopic origin of the
charge density $\rho_e(r)$. We  discuss an explicit example  addressing this question in the next section.
 
\subsection{Equilibrium conditions in general relativity and charge separation}\label{sec:ChargedStar}

\begin{figure}[!htb!]
\begin{center}
	\includegraphics[width=.6\textwidth]{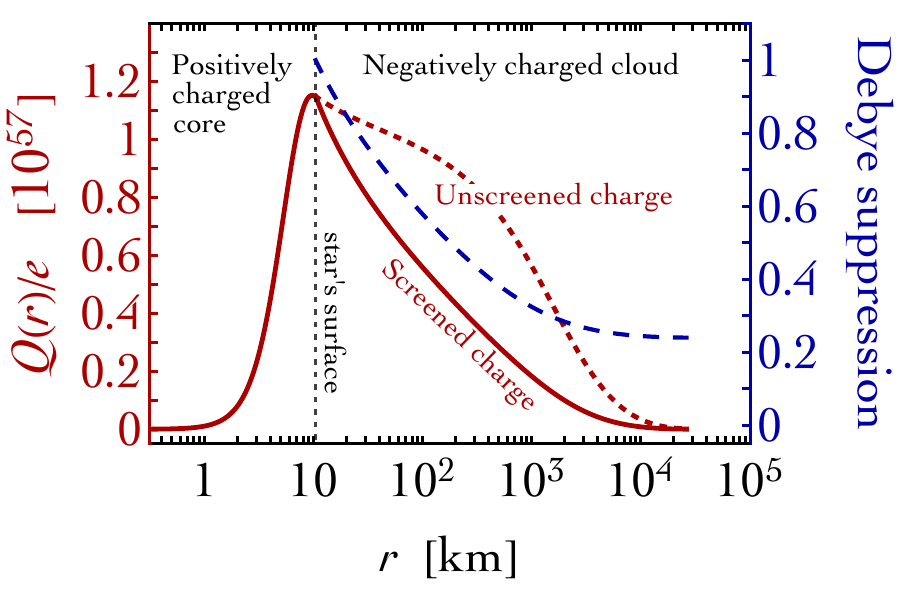}
	\caption{\em \label{fig:ChargeScreen}
Electric charge (in units of the gauge coupling $e$) as a function of the radial distance from the center of the star (dotted red line, left $y$-axes). 
At distances $r>R_{e_-}$  we have $Q(r>R_{e_-}) = 0$ because of charge conservation.
We also plot the Debye suppression factor discussed in eq.~(\ref{eq:DebyeSuppression}) (dashed blue line, right $y$-axes). 
The solid red line is the total charge with the  Debye suppression taken into account, that is $Q(r)s_D(r)$ with $s_D(r)$ the exponential suppression computed in 
eq.~(\ref{eq:DebyeSuppression}).
 }
\end{center}
\end{figure}

We follow the example discussed in ref.~\cite{Ross}, which we consider here in the context of general relativity, and re-elaborate for a generic $U(1)$ dark force.
We refer the interested reader to appendix~\ref{app:Tech} for technical details.  
 The model consists in a star that is made of two charged fluids.
We have a positively charged particle (a \textit{dark proton} with charge $+qe$) with mass $m_{p_+}$ and a negatively charged particle  (a \textit{dark electron} with charge $-qe$) with mass $m_{e_-}$.
We assume $m_{p_+} \gg m_{e_-}$.
In this case, the charge density is 
    \begin{equation}\label{eq:ChargeDensity}
\rho_e(r) = qe\,\left[
n_{p_+}(r) - n_{e_-}(r)
\right]~,~~~~~~e = \sqrt{4\pi\alpha}~,
     \end{equation} 
     where $n_{i}(r)$ is the number density (with $i={p_+},{e_-}$) and $e$ is the gauge coupling. 
     Without loss of generality, we fix $q=1$.
     Furthermore, we assume exact Fermi degeneracy for the two fluids. It means that the number density $n_i$
      is related to the Fermi momentum $p_i$ by $n_i(r) = p_i(r)^3/3\pi^2$.
   \begin{figure}[!htb!]
\begin{center}
$$\includegraphics[width=.4\textwidth]{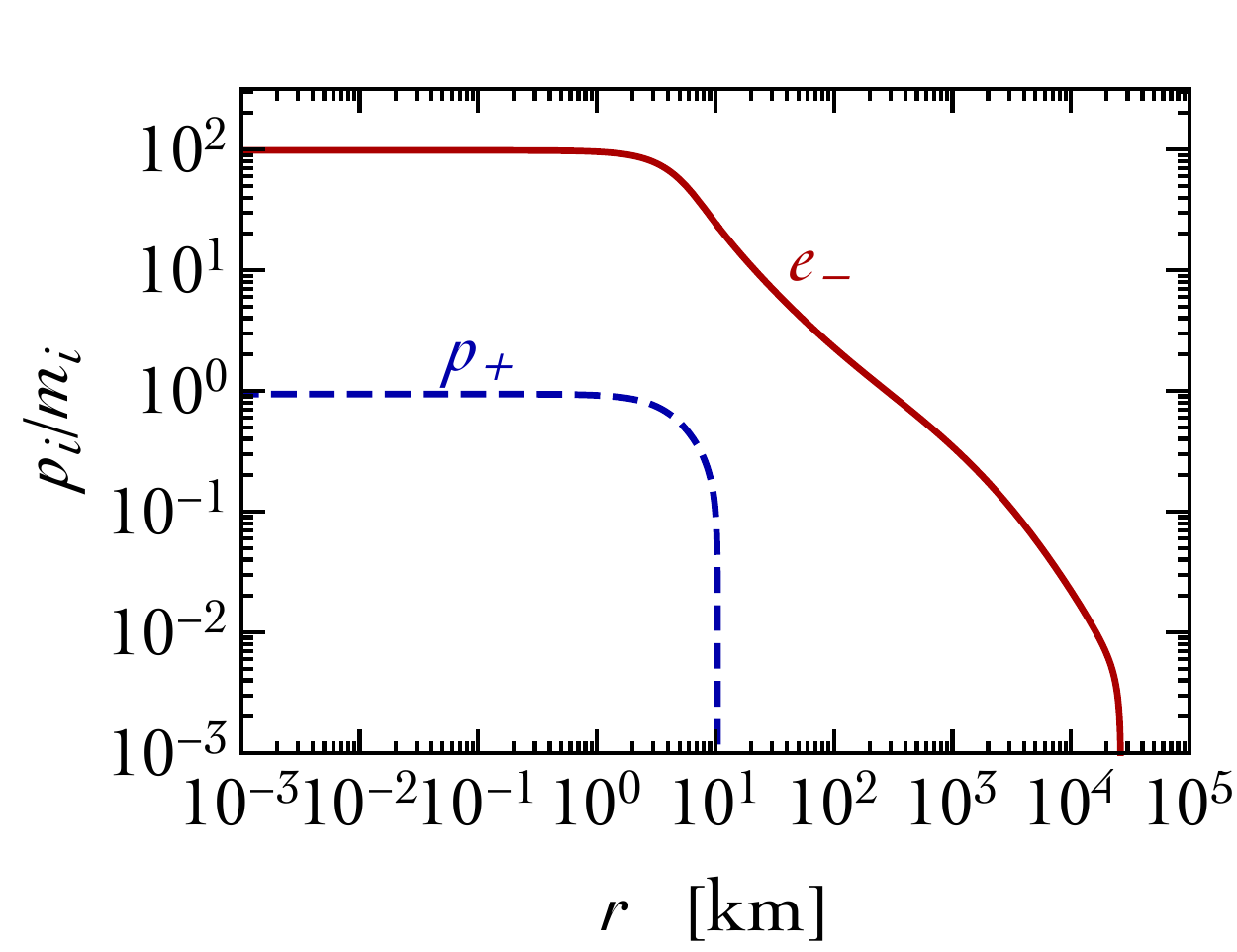}
\qquad\includegraphics[width=.4\textwidth]{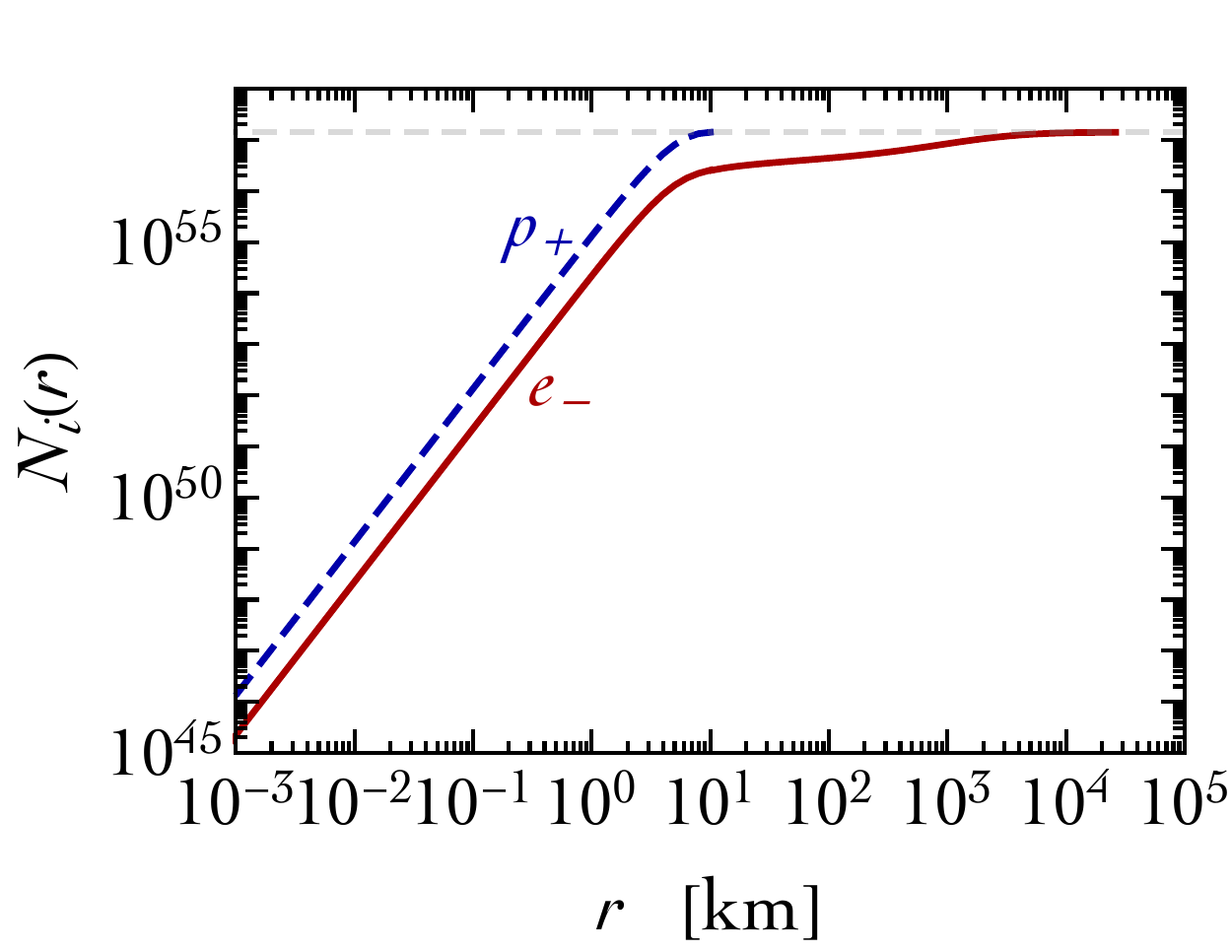}$$\\
\vspace{-1.cm}
$$\includegraphics[width=.4\textwidth]{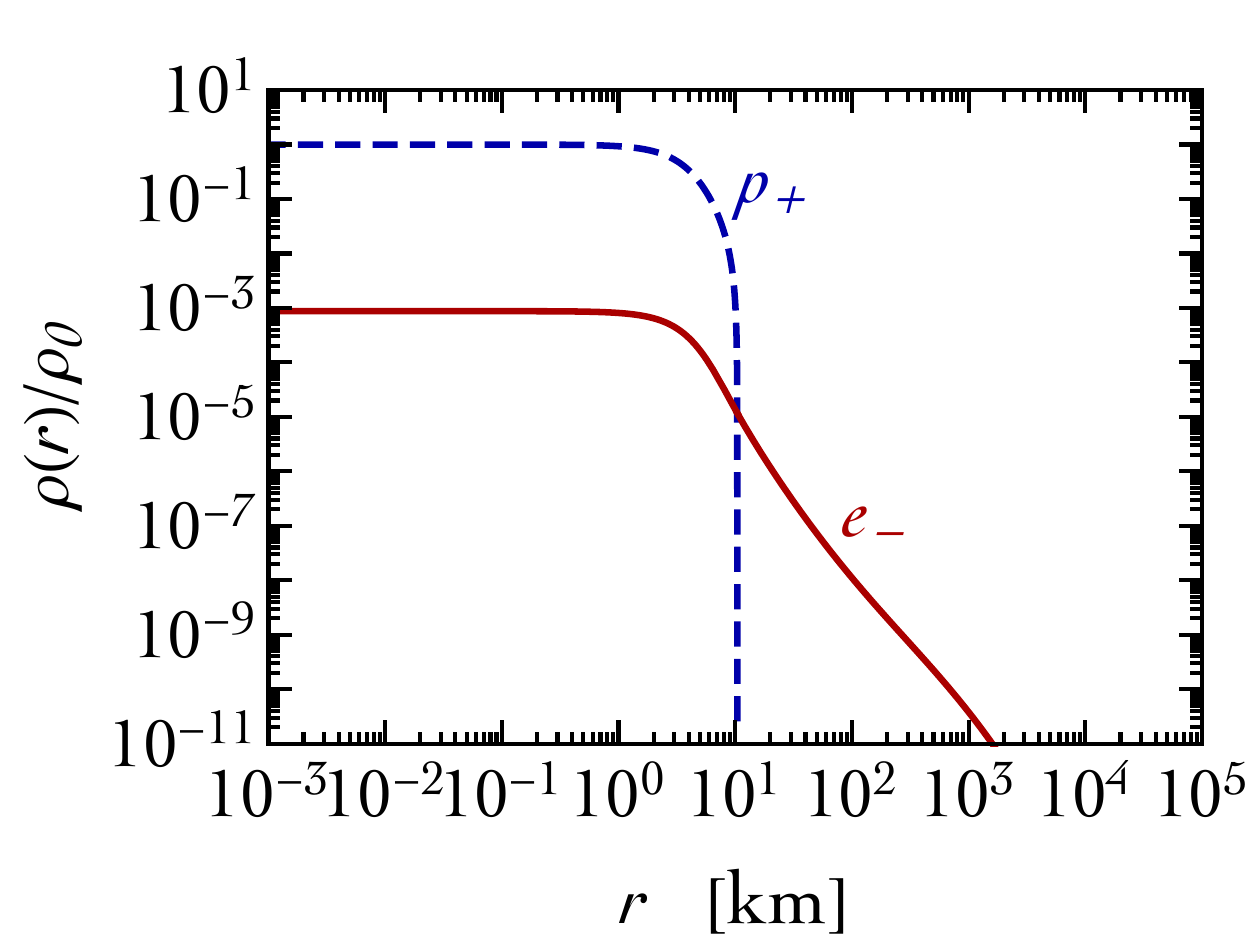}
\qquad\includegraphics[width=.4\textwidth]{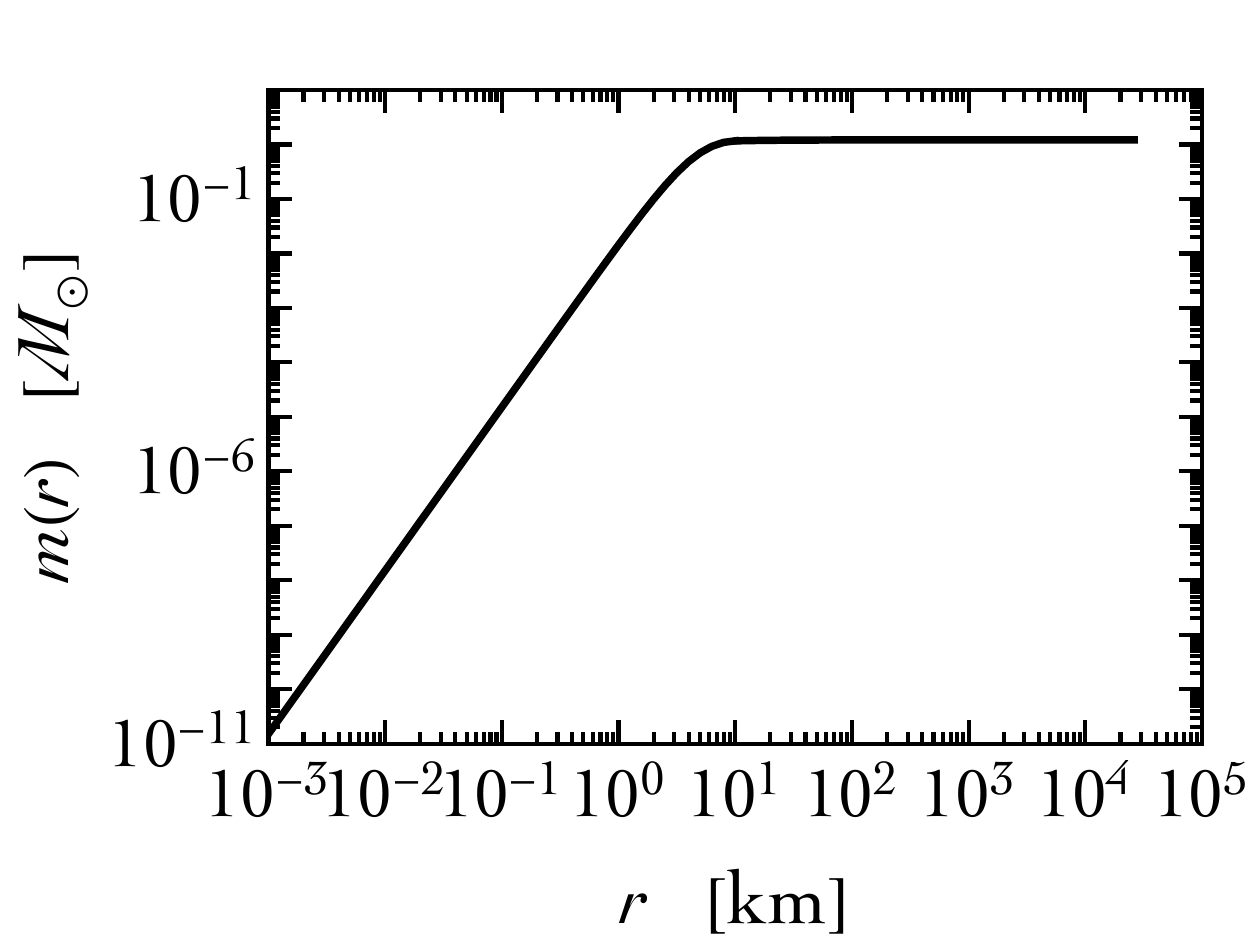}$$
\caption{\em \label{fig:ChargedStar} 
Toy model for charge separation consisting  in a star that is made of two charged fluids for which we assume exact Fermi degeneracy (see section~\ref{sec:ChargedStar}). 
We plot, as a function of the radial distance from the center of the star: in the top-left panel  the distribution of Fermi momenta; in the top-right panel the 
number of particles; in the bottom-left panel the two mass-density distributions normalized to the total value at the center; in the bottom-right panel the mass-energy distribution.
 }
\end{center}
\end{figure}   
     We solve the equilibrium problem in general relativity to find the two density distributions $n_{i}(r)$  (cf.~appendix~\ref{app:Tech}). 
     Consequently, we compute the two radial distances from the center of the star
 $R_i$ defined by $n_{i}(R_i) =0$ (equivalently, 
 $p_{i}(R_i) = 0$). Finally, we  compute,  for each one of the two species,  the number of particles $N_i(r)$  by integrating in space the corresponding number density up to the distance $r$, and we define as total number of particles of type $i$ the quantity $\mathcal{N}_i \equiv N_i(R_i)$. 
     The analysis is based on the following two key points:
     \begin{itemize}
\item [$\circ$]
     We impose the condition of charge conservation. It means that $\mathcal{N}_{p_+} = \mathcal{N}_{e_-}$.
However, we do not impose the condition of charge neutrality, that is $n_{p_+}(r) = n_{e_-}(r)$ (it would imply $\rho_e(r) = 0$ from eq.~(\ref{eq:ChargeDensity})). 
This is becasue we are interested in the two distributions $n_{p_+}(r)$ and $n_{e_-}(r)$ as 
 a result of the equilibrium between gravitational and Coulomb forces.
 We show our results in the top row of fig.\,\ref{fig:ChargedStar}. 
 In the left panel, we plot the two Fermi momenta $p_{p_+}$ and 
 $p_{e_-}$ as function of the radial distance from the center of the star. 
 We see that since $m_{p_+} \gg m_{e_-}$ we have $R_{e_-} \gg R_{p_+}$. This result is intuitively clear: 
 Gravity tends to pack heavier particles closer to the center in comparison with lighter ones.
 This implies that the condition of charge neutrality is not satisfied, since $n_{p_+}(r) \neq n_{e_-}(r)$.
 Nevertheless, charge is conserved. We plot the number of particles $N_i(r)$ in the right panel 
 where one can see that $\mathcal{N}_{p_+} = \mathcal{N}_{e_-}$.
 In our numerics, we consider a specific example that mimics the macroscopic property of a neutron star, that is a
  km-size object with $O(M_{\odot})$ mass. 
  We take $m_{p_+} = 1$ GeV, and $m_{e_-}/m_{p_+} = 5\times 10^{-3}$. 
  Consequently, a solar mass object made of degenerate $p_+$ particles has $\mathcal{N}_{p_+} \approx 10^{57}$.
  \item [$\circ$] The mass of the compact object is primarily 
controlled by the $p_+$ particles (because of $m_{p_+} \gg m_{e_-}$). 
In this case, we have an ambiguity in the definition of the radius of the star since 
    there are two radii $R_{i}$. We proceed in the following way.
    First, we  define the mass density of the star 
\begin{equation}
\rho(r) = \frac{1}{3\pi^2}\left[
m_{p_+} p_{p_+}(r)^3 + m_{e_-}p_{e_-}(r)^3 \right]~.
\end{equation}
Second, we define the radius of the star $R$ to be the distance from the center where the total density
$\rho(r)$ falls to a given fraction $y$ of the value of the density at the center, $\rho_0$. The 
exact value of $y$ can be chosen so that  a further decreasing of $y$ does not alter the mass-energy $m(r)$. 

 We illustrate this procedure in the bottom row of fig.\,\ref{fig:ChargedStar}.
 In the left panel, we plot the ratio $\rho(r)/\rho_0$. 
 The take-home message of this plot is that at radial distances $r\gtrsim R_{p_+}$ 
 the mass density of the star drops by many order of magnitudes, because the only contribution in this region comes from 
 the very light $e_-$ particles.  
 In the right panel, we compute the mass-energy $m(r)$. It is important to remember that in general relativity
 the mass of the star is due to the total contribution of the energy density of the matter and the 
electric energy density, see eq.~(\ref{eq:ChargedMass}) in appendix~\ref{app:Tech}.
 We see that at radial distances $r\geqslant R_{p_+}$ the mass-energy $m(r)$ stays approximatively constant.
It makes sense, therefore, to define the radius of the star as $R \simeq R_{p_+}$ and its mass $M = m(R)$.
 \end{itemize} 
 The charge within the radius $r$ can be obtained by  computing the volume integral of the charge density in eq.~(\ref{eq:ChargeDensity}) (see eq.~(\ref{eq:TotalCharge}) in appendix~\ref{app:Tech}). Because of charge conservation, we have $Q(R_{e_-}) = 0$.
However, $Q(r)\neq 0$ if $r < R_{e_-}$ because of $n_{p_+}(r) \neq n_{e_-}(r)$.
As a consequence of our definition of $R$, therefore, we have 
$\tilde{\mathcal{Q}} = Q(R)\neq 0$, and the outcome of the computation is a compact object with a 
net electric charge. More precisely, the equilibrium solution consists in a positively charged star surrounded by a thin atmosphere of negatively charged particles. The latter extends way further than the actual size of the core (see fig.\,\ref{fig:ChargedStar}). 
Observed from far away, the star appears to be neutral since the negatively charged atmosphere screens the positive core.
At short distance, a net charge emerges.  
We plot the charge $Q(r)/e$ in fig.\,\ref{fig:ChargeScreen} as a function of the radial distance (red lines).
In the numerical example discussed in fig.\,\ref{fig:ChargedStar}, we find
\begin{equation}\label{eq:ChargedValues}
y_{Q} \simeq 0.9~,~~~~\alpha \simeq 5\times 10^{-39}~,
\end{equation}
 which confirms that it is possible to have $y_Q \sim O(1)$ only for super-weak coupling, as discussed in section~\ref{sec:DarkStarsEq}. 
 We checked that the qualitative picture presented in fig.\,\ref{fig:StarPlot} remains true when general relativity is included, as already suggested by the numerical example discussed in this section.

\subsection{Debye length and screening}\label{sec:DarkStarsDebye}

The charging of an astronomical object does not automatically mean that it becomes macroscopically charged at all distances; it depends on the amount of screening.  This can be discussed in terms of a charged object in a plasma. The screening is controlled by the the  plasma Debye length (in natural units)
\be
\lambda_D = \left( \frac{T}{4 \pi \alpha\, n} \right)^{1/2} \, ,
\ee
where  $T$ is the temperature and $n$ the number density of the particles in the plasma. The amount of screening  depends on the  characteristic Debye length of the plasma  compared with the size of the astronomical body. For distances larger of $\lambda_D$, the plasma is neutral and all charged screened; for distances smaller than $\lambda_D$, the plasma is not locally neutral and the screening ineffective.

Because of the super-weak nature of the dark force, the Debye length  is  large for any conceivable scenario.  
For a typical halo temperature $T= 10^6$ K, we find
\be\label{eq:DebyeL}
\lambda_D = 10^{-5} \, 
\left[\left(
\frac{{\rm GeV}^3}{n}\right)
\left(
\frac{10^{-36}}{\alpha}
\right)
\right]^{1/2}
\,{\rm km}~,
\ee
which shows that  for a plasma made  of the  relic density (for DM of mass 10 TeV, $n\simeq 10^{-42}\,{\rm GeV}^3$ ) $\lambda_D \simeq 10^{16}$ km for $\alpha = 10^{-36}$.

We can  estimate the screening produced by the negatively charged cloud in the example discussed in section~\ref{sec:ChargedStar}.
To compute the Debye length in eq.~(\ref{eq:DebyeL}), we use the Fermi energy of the negative charges as an estimate of the temperature. We find (using the number density of a degenerate Fermi gas)
\begin{equation}
\lambda_D^{2} = \frac{3\pi}{4\alpha m_{e_-}^2}\frac{\sqrt{x_{e_-}^2 + 1}}{x_{e_-}^3}~,~~~x_{e_-} \equiv \frac{p_{e_-}}{m_{e_-}}~. 
\end{equation}
For a crude estimate, we can take $x_{e_-}\simeq 1$ (see top-left panel of fig.\,\ref{fig:ChargedStar}), and we find  
$\lambda_D \sim 10^{3}$ km (with the numerical values $\alpha = 5\times 10^{-39}$ and $m_{e_-} = 5\times 10^{-3}$ GeV).
More precisely, we note that the Debye length depends on $r$ by means of  the radial dependence of the number density. 
Using the numerical results in fig.\,\ref{fig:ChargedStar}, we can compute the Debye suppression factor
\begin{equation}\label{eq:DebyeSuppression}
s_D(r) = 
\exp\left[
-\int_{r_0}^r\frac{dt}{\lambda_D(t)}
\right]~,
\end{equation}
with $r_0 = R \simeq R_{p_+}$ the radius of the star, 
and estimate an effective Debye length $\lambda_D^{\rm eff}$ at one e-folding distance, $s_D(\lambda_D^{\rm eff}) = 1/e \simeq 0.37$.
We plot the Debye suppression factor in fig.\,\ref{fig:ChargeScreen} (dashed blue line) which confirms the validity of the previous estimate. 
At distance larger than $\sim 10^3\div 10^4$ km, the star can be considered neutral.

In the presence of a plasma the electrostatic potential generated at distance $r$ by a point-like charge 
$e\mathcal{Q}$, 
$\phi(r) = e\mathcal{Q}/4\pi r$, is Debye-screened to
\begin{equation}\label{eq:DebyeDamp}
\phi(r) \to \phi_D(r) =\frac{e\mathcal{Q}}{4\pi r}e^{-r/\lambda_D}~.
\end{equation}
The screening effect is qualitatively equivalent to the case of a massive dark force mediator, with the Debye length playing the role of the inverse dark photon mass.

In conclusion, 
even if the unbroken $U(1)$ gauge symmetry mediates a long-range force, in realistic situations of physical interest, the force is always screened at distances 
larger than the Debye length.
In the specific example studied in section~\ref{sec:ChargedStar}, the screening effect is provided by the negatively charged particles surrounding the positively charged star. 


\section{Detecting dark forces with gravitational waves}\label{sec:GW}

The coalescence of a compact binary system (made of two black holes, two neutron stars or one black hole and one neutron star) 
can be  divided into three successive stages: the \textit{inspiral}, the \textit{merger} and the \textit{ringdown} phase. 
In this work, we are interested in the physics of the inspiral phase.
As the system evolves during the inspiral phase, it loses energy in the form of gravitational waves. Consequently, 
the two compact objects  are driven closer and closer, and the orbital frequency increases as does the frequency of the emitted gravitational waves and the relative velocity that at the end of the inspiral phase approaches the speed of light.

At the qualitative level, Newtonian dynamics and the  quadrupole formula of general relativity are enough to capture the physics of the inspiral phase.
For point masses $M_1$, $M_2$ in circular orbit, the Kepler's third law relates 
the orbital frequency $\omega$ and the relative radial distance $r$, $\omega^2 = G_N(M_1 +M_2)/r^3$.
The total energy of the binary system $E_{\rm tot}$ is the sum of the kinetic and gravitational potential energy, 
and the power emitted in gravitational wave is  
\be
\mathcal{P}_{\rm GW} = \frac{32}{5} G_N \mu^2 \omega^6 r^4, 
\ee
where $\mu \equiv M_1M_2/(M_1 + M_2)$ is the reduced mass of the system. 
The energy balance equation $dE_{\rm tot}/dt = - \mathcal{P}_{\rm GW}$ can be recast in the form
\begin{equation}\label{eq:PureGravity}
\frac{d\omega}{dt} = \frac{96}{5}\left(
G_N M_C
\right)^{5/3}\omega^{11/3},
\ee
where 
\be
M_C \equiv \mu^{3/5}(M_1 + M_2)^{2/5} = \frac{(M_1 M_2)^{3/5}}{(M_1 + M_2)^{1/5}}\,,
\end{equation}
is the chirp mass of the binary system. 

In binary neutron star mergers, the inspiral phase  of the coalescence that can be observed by the 
advanced LIGO (aLIGO) instrument, takes place when the two 
stars spiral at separation distance 
 in the interval $d_{\rm aLIGO}\sim O(20\div 10^3)$ km. The upper value, $d_{\rm aLIGO}^{\rm \,max} \sim O(10^3)$ km, comes from the minimal frequency detectable by aLIGO, that is $f_{\rm min}\sim O(10)$ Hz. The lower limit, $d_{\rm aLIGO}^{\rm \,min} \sim O(20)$ km, comes from the separation distance at the end of the inspiral phase which is set by twice the typical neutron star radius.
 
\subsection{Dipole radiation}\label{sec:DarkDipole}

We assume that $\lambda_D \gtrsim d_{\rm aLIGO}^{\rm \,max}$. As discussed in section~\ref{sec:DarkStarsDebye}, this assumption corresponds to the case in which 
the charges of the neutron stars are not screened, and we can in first approximation neglect the exponential damping in eq.~(\ref{eq:DebyeDamp}).
The  case in which $d_{\rm aLIGO}^{\rm \,min} \lesssim \lambda_D \lesssim d_{\rm aLIGO}^{\rm \,max}$ is qualitatively equivalent---but physically distinct---to the case of a finite-range massive force carrier if one identifies the mass with the inverse of the Debye length. This case has been investigated in \cite{Kopp:2018jom}. In the following we focus on the case $\lambda_D \gtrsim d_{\rm aLIGO}^{\rm \,max}$ to put more emphasis on the role of the dipole radiation.

The presence of a dark force affects the inspiral dynamics in two crucial ways. 
First, it modifies the long-range interaction potential between the two bodies in the binary system. 
For two compact objects with masses $M_1$ and $M_2$ and charges (in units of the gauge coupling $e$) $\mathcal{Q}_1$ and $\mathcal{Q}_2$ at relative radial distance $r$ we have 
\begin{equation}\label{eq:DarkPot}
V(r) = -\frac{G_N M_1 M_2}{r} + \frac{\alpha \mathcal{Q}_1 \mathcal{Q}_2}{r}~.
\end{equation}
Second, the dark radiation---in addition to the gravitational term---carries away energy from the binary system. 
If the charge-to-mass ratio of the two compact objects is different,  dark radiation is already present with a dipole contribution while 
the first non-zero term for gravity is the quadrupole.
The dynamics is described by the energy-conservation
equation 
\begin{equation}\label{eq:EnergyCon}
\frac{dE_{\rm tot}}{dt} = - \mathcal{P}_{\rm GW} - \mathcal{P}_{\rm dark}~,
\end{equation}
with $E_{\rm tot}$ the total energy of the binary system
\begin{equation}\label{eq:EnergyTot}
E_{\rm tot} = -\frac{G_N \mu (M_1 + M_2)}{r}\bigg(
1 - \underbrace{\frac{\alpha \mathcal{Q}_1 \mathcal{Q}_2}{G_N M_1 M_2}}_{\equiv \tilde{\alpha}}
\bigg)  + \frac{1}{2}\mu r^2 \omega^2.
\end{equation}

The parameter $\tilde{\alpha}$ in \eq{eq:EnergyTot} measures the relative strength of gravity compared with the dark force at the macroscopic level and it is the same defined by \eq{eq:Alpha}.
If  $\tilde{\alpha} > 1$, that is if gravity is weaker, the gravitational potential energy changes sign and the system is not gravitationally bound. 
In order for the binary system to exist, gravity has to be the strongest force, and $\tilde{\alpha} < 1$.

The power emitted in dipole dark radiation is
\begin{equation}
\mathcal{P}_{\rm dark} = \frac{2\alpha \gamma^2 \omega^4 r^2}{3}, \quad \mbox{with} \quad \gamma = \mu\left|
\frac{\mathcal{Q}_1}{M_1} - \frac{\mathcal{Q}_2}{M_2}
\right| .
\end{equation}
The time-dependence of the orbital frequency is governed by the differential equation
\begin{equation}\label{eq:MasterFreq}
\frac{d\omega}{dt} = \frac{96}{5}\left(
G_N M_C
\right)^{5/3}\omega^{11/3}
\bigg(\underbrace{
1 - \frac{\alpha \mathcal{Q}_1 \mathcal{Q}_2}{G_NM_1 M_2}
}_{\rm modified\,quadrupole} \bigg)^{2/3}+ 
\underbrace{2 \mu\omega^3\alpha\left|\frac{\mathcal{Q}_1}{M_1} - \frac{\mathcal{Q}_2}{M_2}\right|^2}_{\rm dipole} ,
\end{equation}
which generalizes the pure gravity result in eq.~(\ref{eq:PureGravity}).
We  solve eq.~(\ref{eq:MasterFreq}) with initial condition at $t=0$ set by 
the minimal frequency detectable by aLIGO,  $f_{\rm min}\sim O(10)$ Hz.

In eq.~(\ref{eq:MasterFreq}), the correction on the quadrupole formula can be mimicked by rescaling the 
chirp mass according to
$M_C \to \tilde{M}_C \equiv M_C\left(
1-\tilde{\alpha}
\right)^{2/5}$. 
This degeneracy is lifted once the dipole radiation is included since it introduces a different $\omega$-scaling if compared to the quadrupole term. 
One can try to absorb the quadrupole correction with a different chirp mass  $\tilde{M}_C$ but imposing the condition $M_1 = M_2$ so that 
the dipole emission is forbidden. 
In this particular situation the effect of dark radiation would be indistinguishable  from pure gravity at least at the leading Newtonian order
(post-Newtonian corrections introduce an explicit dependence from both $M_1$ and $M_2$ thus eliminating, in principle, the freedom 
 to choose  their value to reabsorb the shift $M_C \to \tilde{M}_C$  in the chirp mass). 
 
\begin{figure}[!htb!]
\begin{center}
$$\includegraphics[width=.42\textwidth]{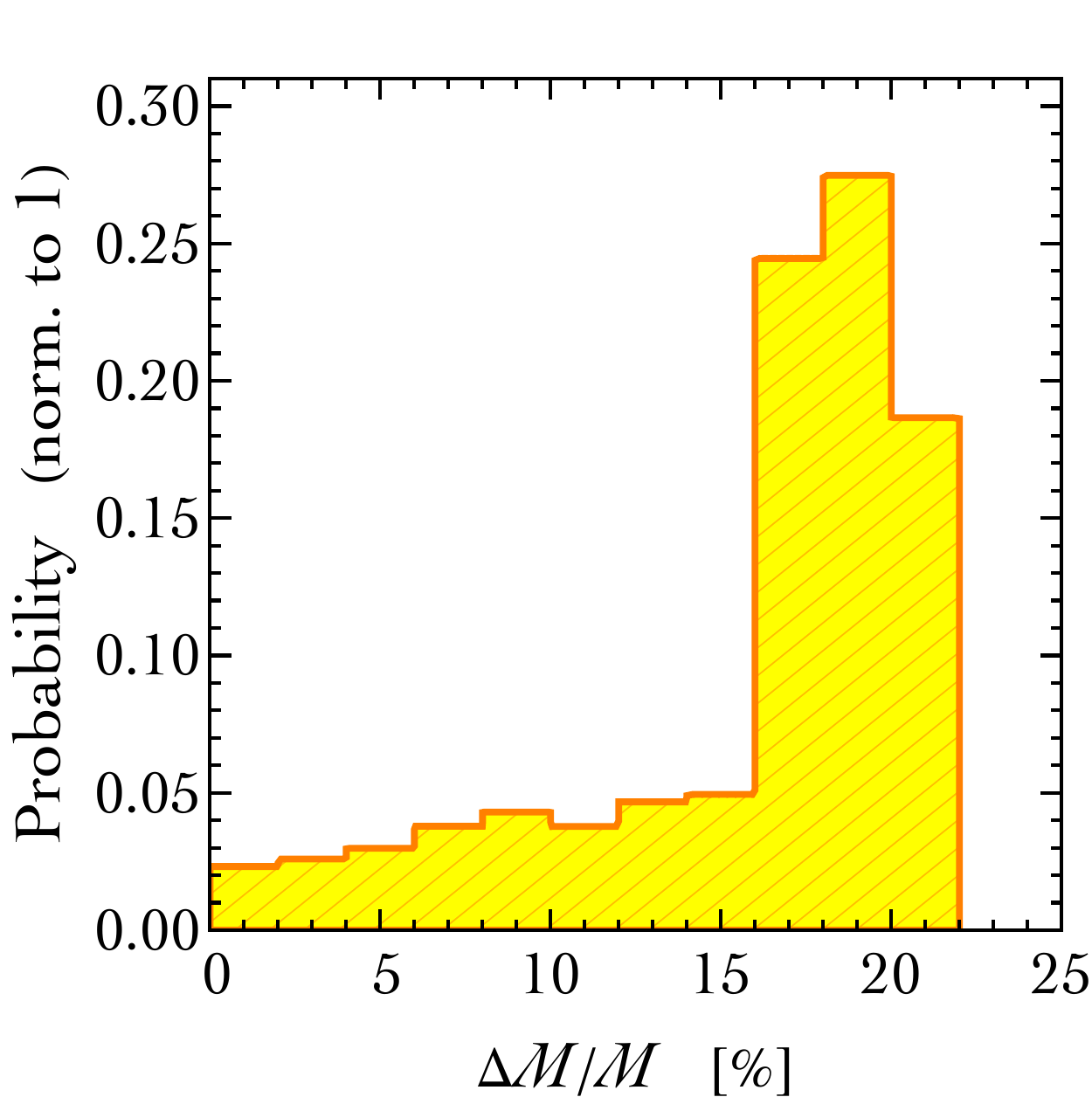}
\qquad\includegraphics[width=.40\textwidth]{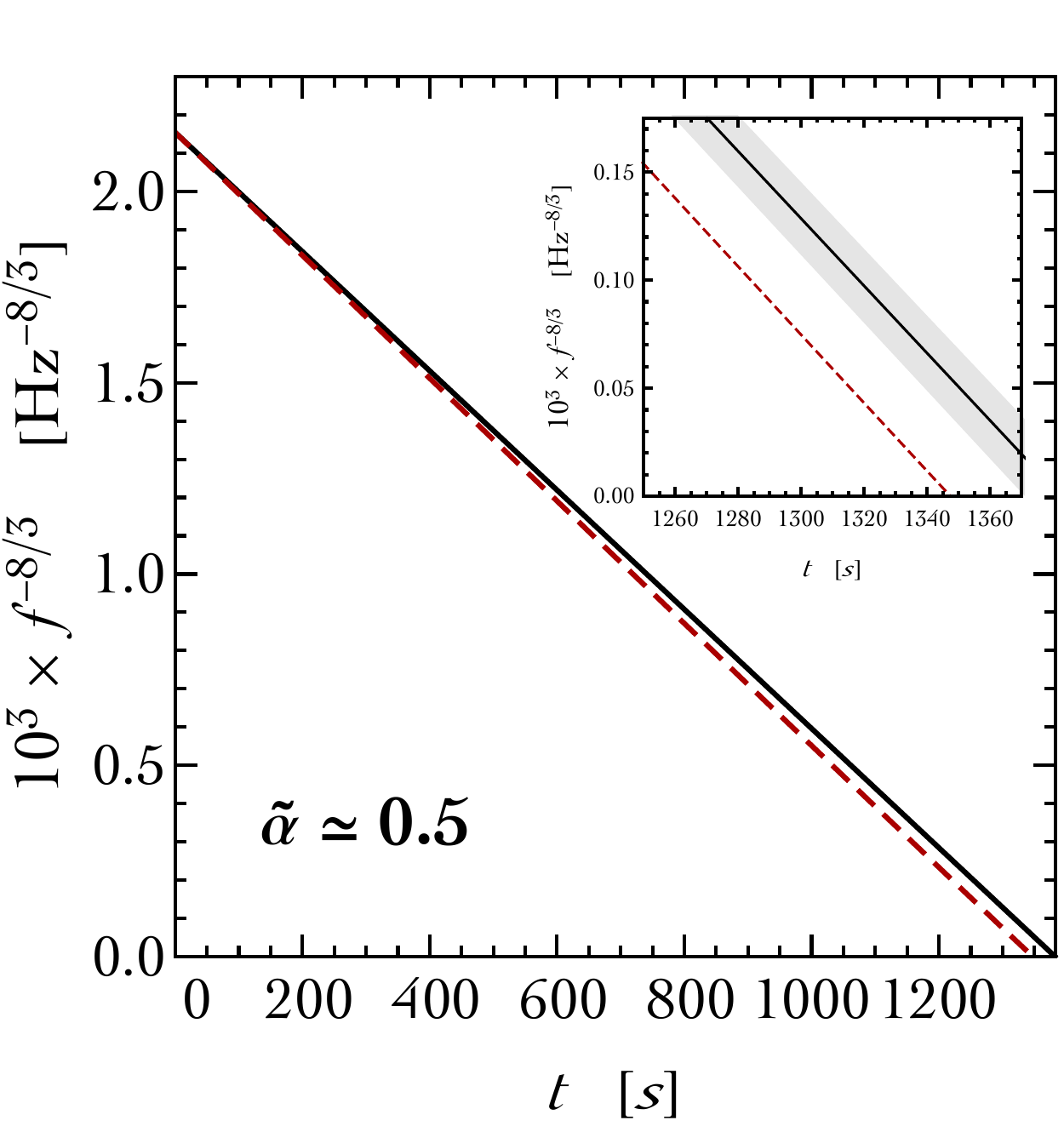}$$
\caption{\em \label{fig:Chirping} 
Left panel. Probability distribution of relative mass difference in binary neutron star systems obtained from the numerical simulations in ref.~\cite{syntheticuniverse}. 
Right panel. Frequency evolution during the inspiral phase in eq.~(\ref{eq:MasterFreq}) (here $f(t) = \omega(t)/\pi$). We plot the case governed by pure gravity (solid black line) and the one with the inclusion of the long-range dark force (dashed red line). The different $\omega$-scaling of the frequency evolution due to the dipole radiation is evident. In the inset, we highlight the small error due to the precise determination of the chirp mass (gray band). 
 }
\end{center}
\end{figure}

 The condition $M_1 = M_2$, however,  seems quite unnatural. 
 To illustrate this point, we analyze the numerical simulation in ref.~\cite{syntheticuniverse} to infer the mass
distribution in binary neutron star systems. We 
show in the left panel of fig.\,\ref{fig:Chirping} our result for the distribution of the mass difference
 $\Delta M/M \equiv |M_1 - M_2|/M_1$. We see that a $O(20\%)$ difference between two solar-mass neutron stars is, most likely,  to be expected.

An explicit example makes more tangible the observable consequences of eq.~(\ref{eq:MasterFreq}).
We assume equal charge, $\mathcal{Q}_1 = \mathcal{Q}_2 \equiv \mathcal{Q}$.
In this case  both the quadrupole and the dipole  terms are controlled by the same quantity $\alpha \mathcal{Q}^2$.
We consider the merger of two neutron star with masses $M_1 \simeq 1.5\,M_{\odot}$ and $M_2 = M_1(1 + 5\%)$, and we take 
$\tilde{\alpha} \simeq 0.5$. In this case the modified quadrupole term can be mimicked in pure GR if one takes $\tilde{M}_C \simeq 1\,M_{\odot}$.
We show in the right panel of fig.\,\ref{fig:Chirping} the evolution of $d\omega/dt$ for these two cases: pure gravity (solid black)
with chirp mass $\tilde{M}_C $ and the case with dark radiation (dashed red).

We find that the dipole correction leads to observable differences (we include a nominal 5 per mill error in $\tilde{M}_C$, see the inset plot).
By decreasing the value of $\tilde{\alpha}$, the dashed red line gets closer to the pure gravity prediction up to the point at which it is no longer possible to distinguish it from the error on the chirp mass. In the numerical case analyzed here, we find that a correction  with
 $\tilde{\alpha}  \gtrsim 0.2$ produces observable effects within a 5 per mill error on the chirp mass measurement, $\Delta M_C/M_C = 0.5\%$. 
 The lower bound on $\tilde{\alpha}$ that can be reached strongly depends on the error on the chirp mass. For instance, if we assume 
 $\Delta M_C/M_C = 0.1\%$  we get $\tilde{\alpha} \gtrsim 0.05$.
 \begin{figure}[!htb!]
\begin{center}
$$\includegraphics[width=.48\textwidth]{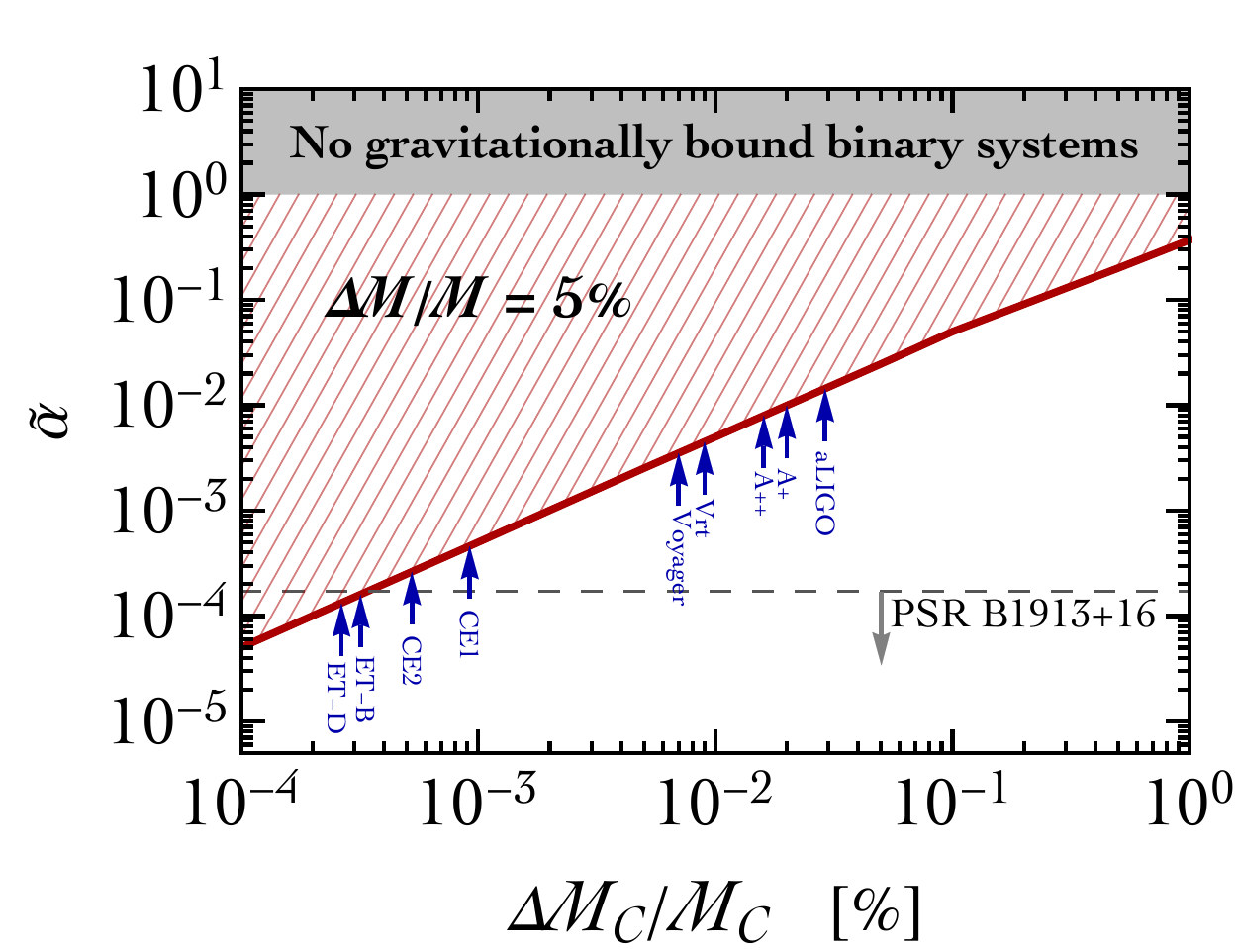}
\qquad\includegraphics[width=.47\textwidth]{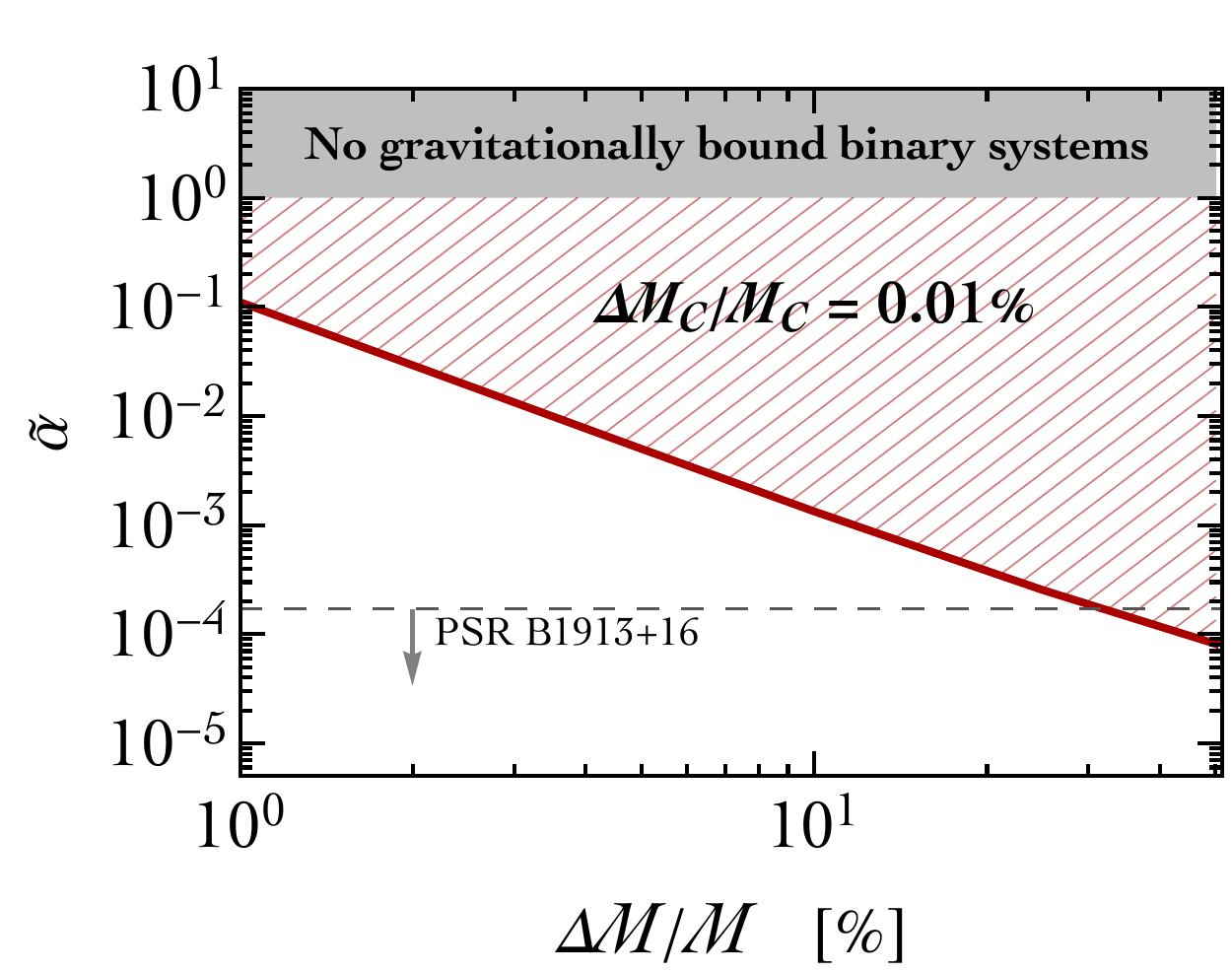}$$
\caption{\em \label{fig:ParameterSpace} 
Detection prospect for $\tilde{\alpha}$ (hatched region with diagonal red lines). 
In the left panel we show this region as a function of the error on the chirp mass  for a fixed value of the relative mass difference between the two neutron stars in the binary system.  In the right panel we show the opposite situation.  The gray region on the top corresponds to values $\tilde{\alpha} > 1$ for which it is not possible to form gravitationally bounded binary systems  since the repulsive force is stronger than gravity and the two stars repel each other.
 The region above the dashed line is excluded by the bound from the Hulse-Taylor binary pulsar PSR B1913+16. The applicability of this bound is discussed in section~\ref{sec:Pulsar}.
In the left panel the small blue arrows mark the sensitivities of present and next-generation gravitational wave interferometers: aLIGO~\cite{TheLIGOScientific:2017qsa,Aasi:2013wya}, various upgrades of aLIGO (A+/A++~\cite{white}, VRT~\cite{white,Adhikari:2013kya}, Voyager~\cite{white}), the Cosmic Explorer  (CE1 and CE2)~\cite{white} and the Einstein Telescope (ET-B and ET-D)~\cite{avi,Hild:2009ns}.
 }
\end{center}
\end{figure}

The detection prospect is summarized in fig.\,\ref{fig:ParameterSpace} where the region shaded in red represents the values of $\tilde{\alpha}$ that could be probed 
 by gravitational wave interferometers.
In the left panel, we fix the
mass ratio of the inspiralling neutron stars ($\Delta M/M = 5\%$) and we show the detection prospect 
as a function of the error on the chirp mass. In the right panel, we fix the latter 
($\Delta M_C/M_C = 0.01\%$) and we take the mass ratio $\Delta M/M$ as a free parameter (fig.\,\ref{fig:Chirping}, left panel for the expected mass distribution of neutron star binaries in the local Universe). 
In the same left panel of fig.\,\ref{fig:ParameterSpace} we also mark with vertical blue arrows the expected and projected  sensitivities to the chirp mass in a binary neutron star merger. To determine these sensitivities, we use the results of ref.~\cite{Alexander:2018qzg}, to which we refer for further details.\footnote{In particular, we quote here the sensitivities obtained using the waveforms of general relativity.}
 We stress that the exact value of the error on the chirp mass depends on the luminosity distance of the detected event (in ref.~\cite{Alexander:2018qzg} taken to be $100$ Mpc for the neutron star merger). 
 
 We conclude that gravitational-wave interferometers have the possibility to test the existence of forces weaker than gravity for values $\tilde{\alpha} \gtrsim 10^{-4}\div 10^{-5}$ (subject to the caveats discussed in the text).
 
 Two comments are in order. 
 In eq.~(\ref{eq:MasterFreq}) we  compared the result obtained in pure gravity for $d\omega/dt$  with the correction due to the dark force.
  The crucial point for the relevance of the analysis is that the dipole radiation introduces a different $\omega$-scaling if compared with the quadrupole term. 
  Without this correction, the effect of the dark force can not be observed. 
  A first comment is that this is only true under the assumption that $\lambda_D \gtrsim d_{\rm aLIGO}^{\rm \,max}$. In the situation $d_{\rm aLIGO}^{\rm \,min} \lesssim \lambda_D \lesssim d_{\rm aLIGO}^{\rm \,max}$, the effect of the dark force can be observed even in the absence of dipole radiation since the partial Debye screening turns the long-range Coulomb force into a finite-range interaction, thus introducing an extra $r$-dependence in eq.~(\ref{eq:DarkPot}) of the form
  \begin{equation}\label{eq:DarkPotDeb} 
V(r) = -\frac{G_N M_1 M_2}{r} + \frac{\alpha \mathcal{Q}_1 \mathcal{Q}_2}{r}e^{-r/\lambda_D}~,
\end{equation}
that can not be reabsorbed in a redefinition of the masses.
  A second subtlety is that we derived eq.~(\ref{eq:MasterFreq}) using Newtonian dynamics and  Einstein's quadrupole formula of general relativity for the power emitted in gravitational waves.   This is only the leading order description. 
  Post-Newtonian corrections play a very  important role, especially during the latest stage of the inspiral prior to the actual plunge and merger~\cite{Damour:2009wj}. 
  It is, therefore, legitimate to ask up to what extent the inclusion of post-Newtonian corrections may generate some degeneracy with the $\omega^3$-dependence of dipole radiation in eq.~(\ref{eq:MasterFreq}). 
  We address this question in appendix~\ref{app:PN} in which we include post-Newtonian corrections up to the 3.5\,PN order. We argue that the presence of 
  post-Newtonian corrections does not introduce any degeneracy. On the contrary, we find that including post-Newtonian corrections exacerbates the difference between the case with and without dark radiation. 

\subsection{Post-Keplerian pulsar timing parameters}\label{sec:Pulsar}

In this section we compare the bound on $\tilde{\alpha}$ computed in the previous section with those that can be extracted from 
binary pulsars. 
A binary pulsar is a binary system 
formed by a pulsar (that is a highly magnetized rotating neutron star emitting a beam of electromagnetic radiation) with a companion, usually a white dwarf or another neutron star.  The orbital period of the binary system decreases as a consequence of energy loss due to the emission of 
gravitational waves.
The measurement of the rate of change of the orbital period, therefore, provides a powerful test of general relativity. 

From the raw data of binary pulsars it is possible to extract with very good precision three post-Keplerian observables: the secular change of the orbital 
period, $\dot{P_b}$, the secular rate of advance of the periastron, $\dot{\omega}$, and the gravitational redshift, $\gamma$.  
Notice that, on the contrary, the masses of the two stars are \textit{a priori }unknown.
Consequently, one can compare the theoretical predictions for the three observables $\dot{P_b}$, $\dot{\omega}$, $\gamma$ with their measured values as a function of the unknown masses $M_1$, $M_2$. 

 \begin{figure}[!htb!]
\begin{center}
$$\includegraphics[width=.45\textwidth]{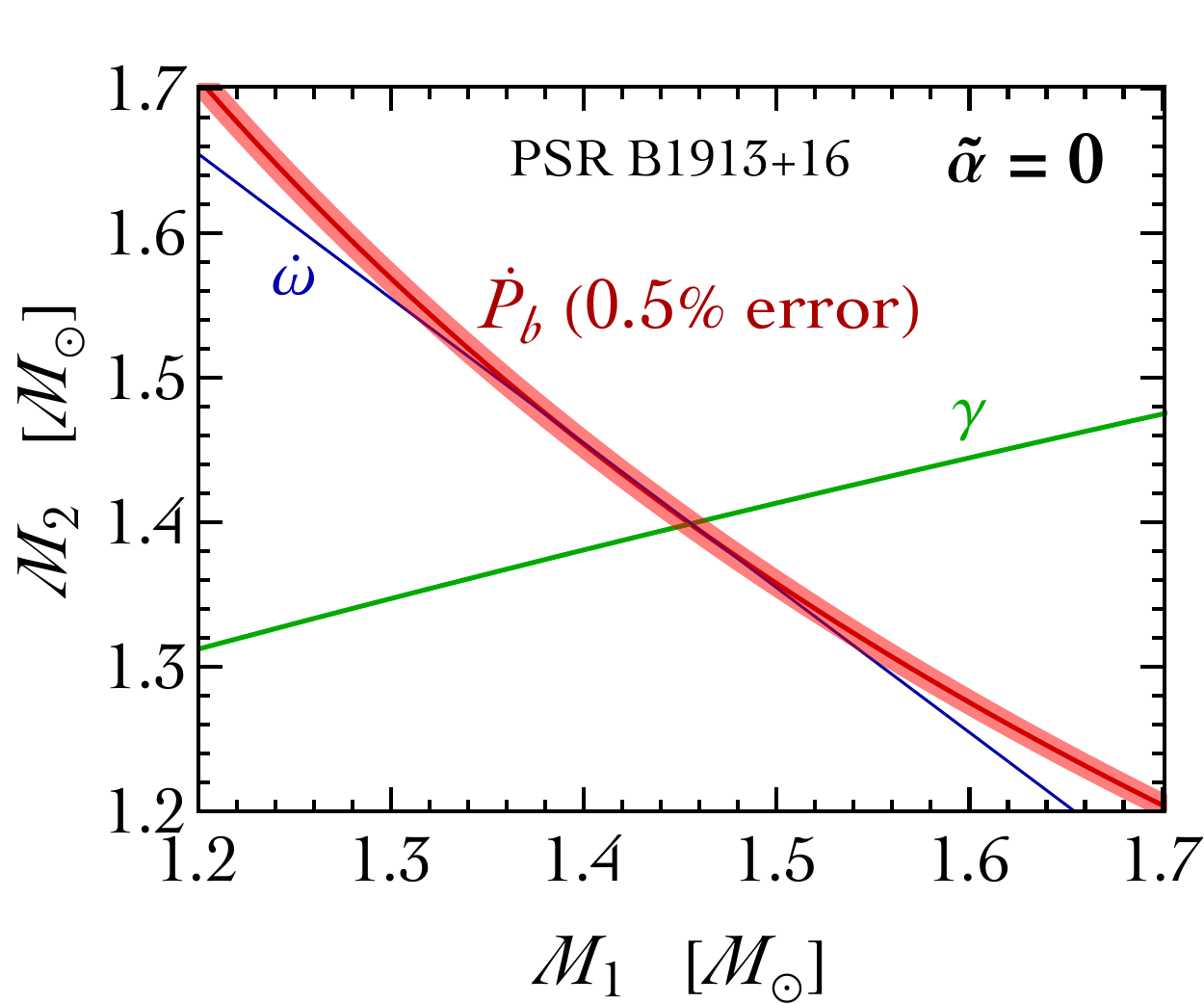}
\qquad\includegraphics[width=.45\textwidth]{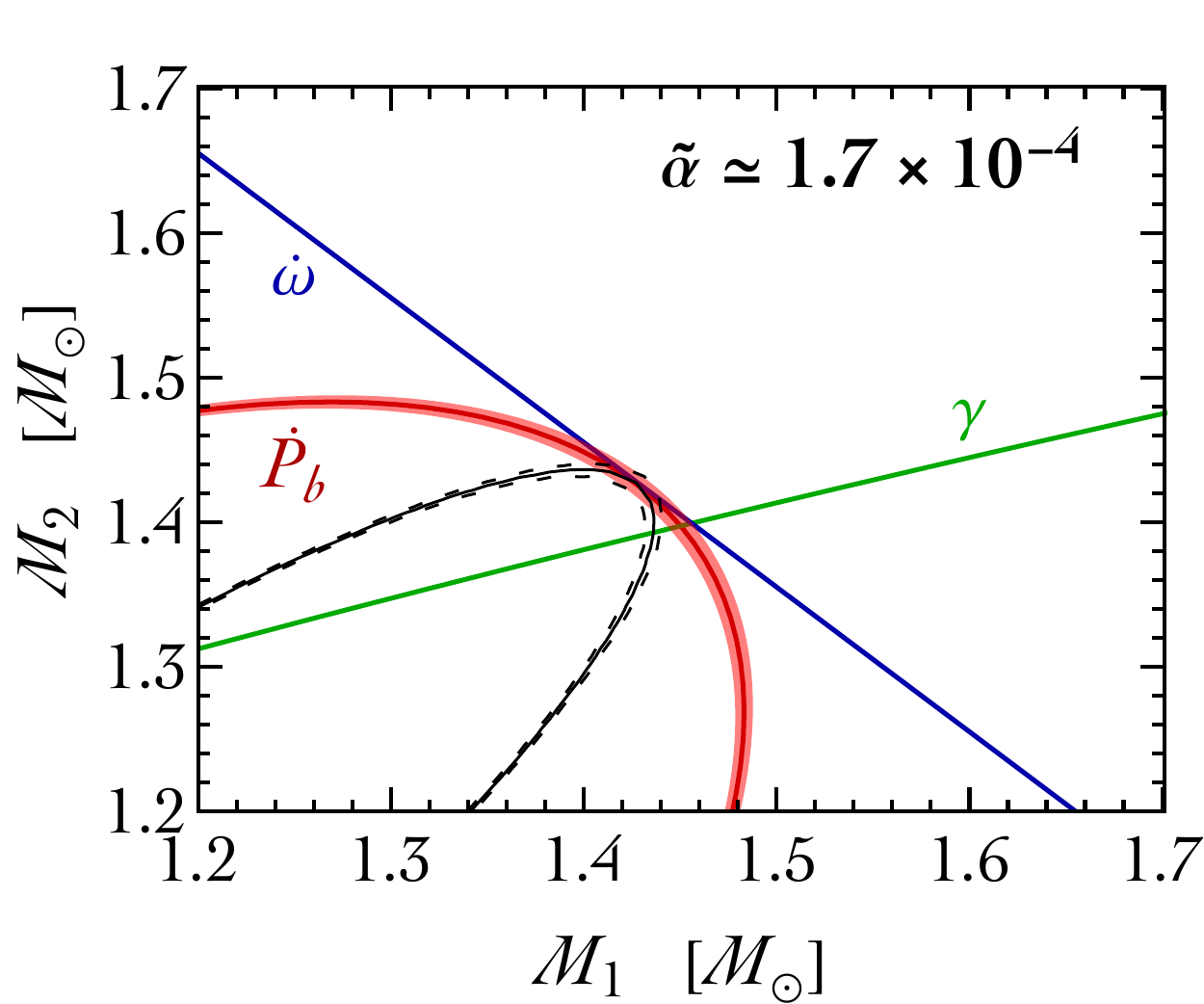}$$
\caption{\em \label{fig:Pulsar} 
The $\gamma$-$\dot{\omega}$-$\dot{P}$ test for the Hulse-Taylor binary pulsar PSR B1913+16 in general relativity 
(left panel) and in the presence of an additional dark force (right panel).
We include a $0.5\%$ error on the measurement of the 
secular change of the orbital 
period $\dot{P}_b$ (red band).
The errors on the periastron advance rate $\dot{\omega}$ and on the gravitational redshift $\gamma$ are comparable with the
 thickness of the corresponding lines.
 }
\end{center}
\end{figure}

For the binary pulsar to exist, there must be a point in the mass plane $(M_1,M_2)$ where the three lines meet each other (dubbed the ``$\gamma$-$\dot{\omega}$-$\dot{P}$ test'').
For definiteness, let us focus on the Hulse-Taylor binary (a.k.a. PSR B1913+16). This is a binary system formed by a pulsar and another neutron star.
In addition to being the first binary pulsar discovered, it remains one of the best laboratory for studying general relativity. 
We use the experimental data quoted in ref.~\cite{Weisberg:2004hi}.

The result of the $\gamma$-$\dot{\omega}$-$\dot{P}$ test in general relativity (corresponding to $\tilde{\alpha}=0$) is illustrated in the left panel of fig.\,\ref{fig:Pulsar}. The largest uncertainty is that on  $\dot{P_b}$, and we plot in red a band corresponding to a nominal $0.5\%$ error.\footnote{This large error arises  because the measured value of the orbital period decay, $\dot{P}_{b,{\rm obs}}$, before being compared with the theoretical prediction, must be corrected with a term accounting for the relative acceleration between the solar system and the binary pulsar, $\dot{P}_{b,{\rm Gal}}$. This correction depends on several poorly known quantities (including the distance and proper motion of the pulsar and the radius of the Sun's galactic orbit), and 
this affects the precision of the measurement. Ref.~\cite{Weisberg:2004hi} quotes the corrected value 
$\dot{P}_{b,{\rm obs}} - \dot{P}_{b,{\rm Gal}} = -(2.4056\pm 0.0051)\times 10^{-12}$, corresponding to a $\sim 0.6\%$ error at the $3\sigma$ level. The recent analysis in ref.~\cite{Weisberg:2016jye} quotes the value $\dot{P}_{b,{\rm obs}} - \dot{P}_{b,{\rm Gal}} = -(2.398\pm 0.004)\times 10^{-12}$, corresponding to a $\sim 0.5\%$ error at the $3\sigma$ level.} General relativity passes the test, and it is possible to extract the values of the masses $M_1$, $M_2$ in the binary system.

Following the spirit of this section, we now assume that  both stars in the binary system have a dark charge. The presence of 
an additional dark force modifies the theoretical prediction  of the  three observables $\dot{P_b}$, $\dot{\omega}$, $\gamma$.
Using $\omega = 2\pi/P_b$, we find for the secular change of the orbital 
period
\begin{eqnarray}
\dot{P_b} &=& -\frac{192\pi}{5}\left(
G_N M_C
\right)^{5/3}
\bigg(\underbrace{
1 - \frac{\alpha \mathcal{Q}_1 \mathcal{Q}_2}{G_NM_1 M_2}
}_{\rm modified\,quadrupole} \bigg)^{2/3}
\left(
\frac{2\pi}{P_b}
\right)^{5/3} 
\underbrace{\left(
1 + \frac{73}{24}\mathbf{e}^2 + \frac{37}{96}\mathbf{e}^4
\right)(1-\mathbf{e}^2)^{-7/2}}_{\rm eccentricity}\nonumber\\
&-&\underbrace{4\pi\left(
\frac{2\pi}{P_b}
\right)\mu\alpha\left|\frac{\mathcal{Q}_1}{M_1} - \frac{\mathcal{Q}_2}{M_2}\right|^2}_{\rm dipole}
\underbrace{\frac{1+\mathbf{e}^2/2}{(1-\mathbf{e}^2)^{5/2}}}_{\rm eccentricity}~,
\end{eqnarray}
where $\mathbf{e}$ is the orbital eccentricity. Compared with the pure gravity result, 
we have--in analogy with eq.~(\ref{eq:MasterFreq})--a change in the quadrupole radiation and an additional contribution due to dipole dark 
radiation.
The theoretical prediction for the periastron advance rate is 
\begin{equation}
\dot{\omega} = \frac{3}{1-\mathbf{e}^2}\left(
\frac{2\pi}{P_b}
\right)\left[
G_N(M_1 + M_2)\left(
\frac{2\pi}{P_b}
\right)
\right]^{2/3}\bigg(\underbrace{
1 - \frac{\alpha \mathcal{Q}_1 \mathcal{Q}_2}{G_NM_1 M_2}
}_{{\rm shift\,of\,}G_N} \bigg)^{2/3}~.\label{eq:OmegaTest}
\end{equation}
Similarly, for the gravitational redshift we find
\begin{equation}
\gamma = 
\frac{\mathbf{e}}{(2\pi/P_b)}\left(\frac{M_2}{M_1 + M_2}\right)\left[
G_N(M_1 + M_2)\left(
\frac{2\pi}{P_b}
\right)
\right]^{2/3}
\bigg(\underbrace{
1 - \frac{\alpha \mathcal{Q}_1 \mathcal{Q}_2}{G_NM_1 M_2}
}_{{\rm shift\,of\,}G_N} \bigg)^{2/3}\left(
\frac{M_2}{M_1 + M_2} + 1
\right)~.\label{eq:GammaTest}
\end{equation}
In eq.s~(\ref{eq:OmegaTest},\ref{eq:GammaTest}), the presence of the dark force amounts to a shift in $G_N$ (cf.~eq.~(\ref{eq:DarkPot})) w.r.t. the result in pure gravity.

To simplify the analysis, we assume equal charge, $\mathcal{Q}_1 = \mathcal{Q}_2 \equiv \mathcal{Q}$. 
As in the previous section, corrections w.r.t. general relativity are, therefore, controlled by the parameter $\tilde{\alpha}$.
In the right panel of fig.\,\ref{fig:Pulsar} we show that values 
$\tilde{\alpha} \lesssim  1.7\times 10^{-4}$ are  compatible with the existence of the binary pulsar within the error on $\dot{P_b}$ (for comparison, we plot with dashed black lines the value of $\dot{P_b}$ for $\tilde{\alpha} = 10^{-3}$; clearly, in this case there is no solution to the 
$\gamma$-$\dot{\omega}$-$\dot{P}$ test). 
 
This simple analysis shows that pulsar timing provides a constraint on $\tilde{\alpha}$ that is comparable with  
the values explored by gravitational wave interferometers, as shown in fig.\,\ref{fig:ParameterSpace}. 
If one takes the comparison at face value, it is probably fair to say that pulsar timing measurements  
already rule out values of $\tilde{\alpha}$ that are relevant for aLIGO given its present sensitivity.
Similar conclusions were obtained in the context of scalar-tensor gravity~\cite{Damour:1998jk}. 

 \begin{figure}[!th]
\begin{center}
	\includegraphics[width=.5\textwidth]{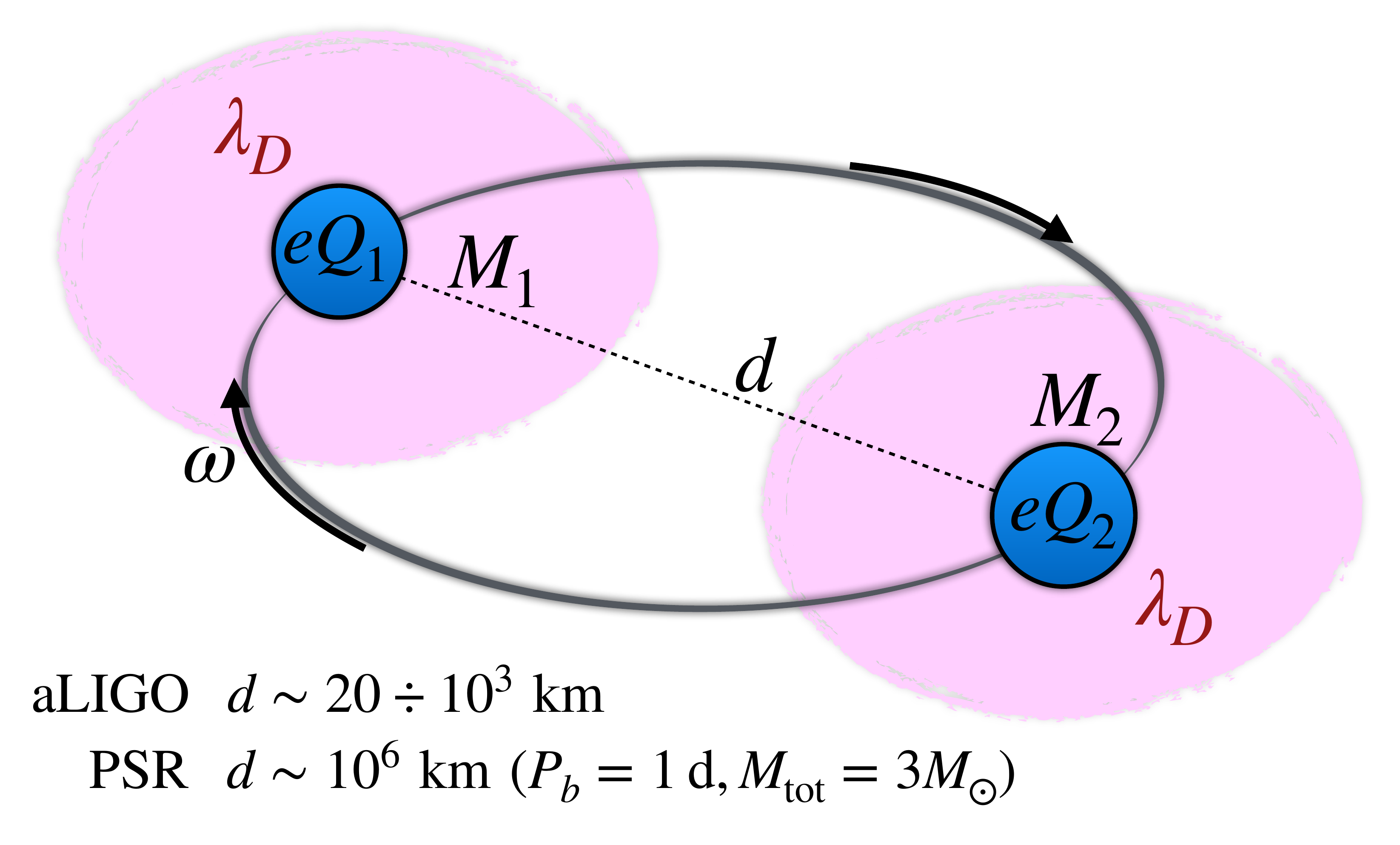}
	\caption{\em \label{fig:Debye}  
	Schematic of a neutron star binary system at typical separation distance $d$ and orbital frequency $\omega$. The two stars with masses $M_1$, $M_2$  have dark charges 
	$e\mathcal{Q}_1$, $e\mathcal{Q}_2$ under an unbroken $U(1)$ symmetry. 
	The Debye length of the surrounding medium is $\lambda_D$. At distace $d\gtrsim \lambda_D$, the
	dark charges are Debye-screened, and the dynamics governed solely by gravitational interactions. 	
 }
\end{center}
\end{figure}

 Yet there is an important caveat that must be considered when comparing pulsar time constraints and 
 detection prospects in gravitational wave interferometers. 

As already discussed, in a binary system the separation distance during the  inspiral phase  of the coalescence ranges in the interval $d_{\rm aLIGO}\sim O(20\div 10^3)$ km. In a binary pulsar, on the contrary, the typical distance between the two stars
is much greater, and for a typical orbital period $P_b = 1$ d and total mass $M_{\rm tot} = 3\,M_{\odot}$
one finds (considering the semi-major axes of the orbit as an order-of-magnitude estimate) $d_{\rm PSR} \sim 10^6$ km.
The two stars are not in the perfect vacuum, and, therefore, it is conceivable to expect that their charges are Debye-screened for distances larger than the 
Debye length $\lambda_D$ of the medium surrounding them. The typical situation is sketched in fig.\,\ref{fig:Debye}.
If the condition $d_{\rm aLIGO}  \lesssim \lambda_D  \lesssim d_{\rm PSR}$ occurs, 
 the dark force does not alter the binary pulsar but it becomes relevant for neutron star mergers.
In such case, the bound on $\tilde{\alpha}$ derived from pulsar timing measurements does not apply. 
This is the situation discussed in section~\ref{sec:DarkStars}.

\section{Neutron stars with a dark charge}\label{sec:ChargedNS}

We now  consider the case of charged neutron stars. At the microscopic level, we need a mechanism to explain the presence of dark sector particles inside a neutron star.  Dark states can be present inside a neutron star because of being there since its formation (in supernovae) or because of a process of accretion during its lifetime. The latter can take place by capture  of dark matter from the surrounding halo or by means of some of the neutrons in the star decaying into the dark sector.

The possibility of producing dark matter in the supernova explosion creating the neutron star seems excluded because the  \textit{bremsstrahlung} process on which this mechanism is based requires the dark force to couple to ordinary matter~\cite{Nelson:2018xtr}---which is not the case we are considering. The capture from the surrounding halo with net final charge---beside requiring  a relic density made of DM of opposite charges and very different masses~\cite{Alexander:2018qzg}---produces a negligible number of captured dark states~\cite{Gould:1987ir,McDermott:2011jp,Nelson:2018xtr,Kopp:2018jom,Alexander:2018qzg}.
This leaves  only the scenario in which dark matter is produced inside the  star by the decay of the neutrons.

At the macroscopic level, the challenge is  to find equilibrium solutions of Einstein-Maxwell field equations that are compatible with the observed properties of neutron stars (as far as mass, radius and tidal deformability are concerned) and that, at the same time, generate a net macroscopic charge. 

As far as the first point is concerned, we focus on the effect on the stellar structure of an invisible neutron decay process into the dark sector~\cite{Fornal:2018eol,Cline:2018ami}. We take, as benchmark case, the decay mode 
\begin{equation}\label{eq:DarkDecay}
n \to  Q_D + Q_D + Q_U~,
\end{equation}
studied in~\cite{Barducci:2018rlx}.
Charge conservation imposes the constraint $2q_{D} + q_{U} = 0$ on the $U(1)$ dark charges of the final state particles.
There are two possibilities to be explored:

\begin{enumerate}

\item[{\it i)}] {\underline{Invisible decay of bound neutrons}}. The absence of an invisible decay channel for the bound 
neutron
in $^{16}$O and $^{12}$C puts a stringent lower limit on the neutron life-time, $\tau_n^{\rm inv} \gtrsim 10^{29}$ years~\cite{Ahmed:2003sy}.  
However, it is still possible to speculate about the existence of an invisible decay channel 
satisfying this bound. In such case, the only mass constraint applicable is the
kinematic threshold
 \begin{equation}
2m_{D} + m_{U} < m_n = 939.565~{\rm MeV}~.
\end{equation}
We have $N_0 \approx 10^{57}$ neutrons in a solar-mass neutron star. In the presence of the invisible decay $n \to Q$,
 after a time period $t$ we have $N_Q(t) = N_0\left[
 1 - \exp\left(-t/\tau_n^{\rm inv}
 \right)\right]$ dark particles inside the neutron star.
Most of the neutron stars are billions of years old.   
This means that in our Universe today it is plausible to observe neutron stars 
hosting approximately 
 \begin{equation}\label{eq:DarkParticle}
 N_Q(t=10^{9}\,{\rm yr}) \approx 10^{9}\frac{N_0}{\tau_n^{\rm inv}} \approx 10^{37}~,
 \end{equation}
 dark particles in their core. In eq.~(\ref{eq:DarkParticle}) we assumed a life-time close to the present 
 bound. We have a charged fraction $y_Q = N_Q/N_0 \sim O(10^{-20})$, which is too small (fig.\,\ref{fig:StarPlot}).
 
\item[{\it ii)}] {\underline{Decay of free neutrons and the neutron life-time puzzle}}.
It is possible to avoid the strong limit on the neutron life-time in bound systems if one takes the
 narrow mass window
\begin{equation}
\underbrace{937.900~{\rm MeV}}_{{\rm decay\,of\,\,}^9{\rm Be\,closed}}<2m_{D} + m_{U} < \underbrace{m_n = 939.565~{\rm MeV}}_{\rm decay\,kin.\,open}
\end{equation}
in which all nuclear decay limits are satisfied. This is a extremely tuned mass range but it is motivated by the 
possibility to solve the neutron life-time puzzle ~\cite{Paul:2009md} (the life-time of the free neutron 
determined with the \textit{beam} method~\cite{Yue:2013qrc} is 8 $s$ longer 
than the one determined by means of the \textit{bottle} method~\cite{Serebrov:2017bzo}). 
In this case we expect an inverse width for the dark decay of the order of the second,  
and it is in principle possible to collect in the neutron star much more dark particles compared to case {\it i)}. 
In the following section we shall focus on case {\it ii)} for our numerical results. 
 
\end{enumerate}

The chemical equilibrium for the reaction in eq.~(\ref{eq:DarkDecay}) is $\mu_n = 2\mu_{D} + \mu_{U}$.
As a consequence, we have only two independent chemical potentials.
In this regime, the neutron decay is equilibrated with the inverse process. 

The neutron star is still neutral because the charge is conserved in the decay process.
In the next section, we shall discuss, along the lines of section~\ref{sec:ChargedStar}, the mechanism of charge separation.

\subsection{Charging a neutron star}\label{sec:ChargedNS}

The electric charge density is given by
\begin{equation}
\rho_e(r) = e\left[q_{D}n_{D}(r) + q_{U}n_{U}(r)\right] = q_{D}e\left[n_{D}(r) - 2n_{U}(r)\right]~,
\end{equation}
where we used conservation of charge. 
Without loss of generality, we take $q_{D} = 1$.
 \begin{figure}[!htb!]
\begin{center}
$$\includegraphics[width=.35\textwidth]{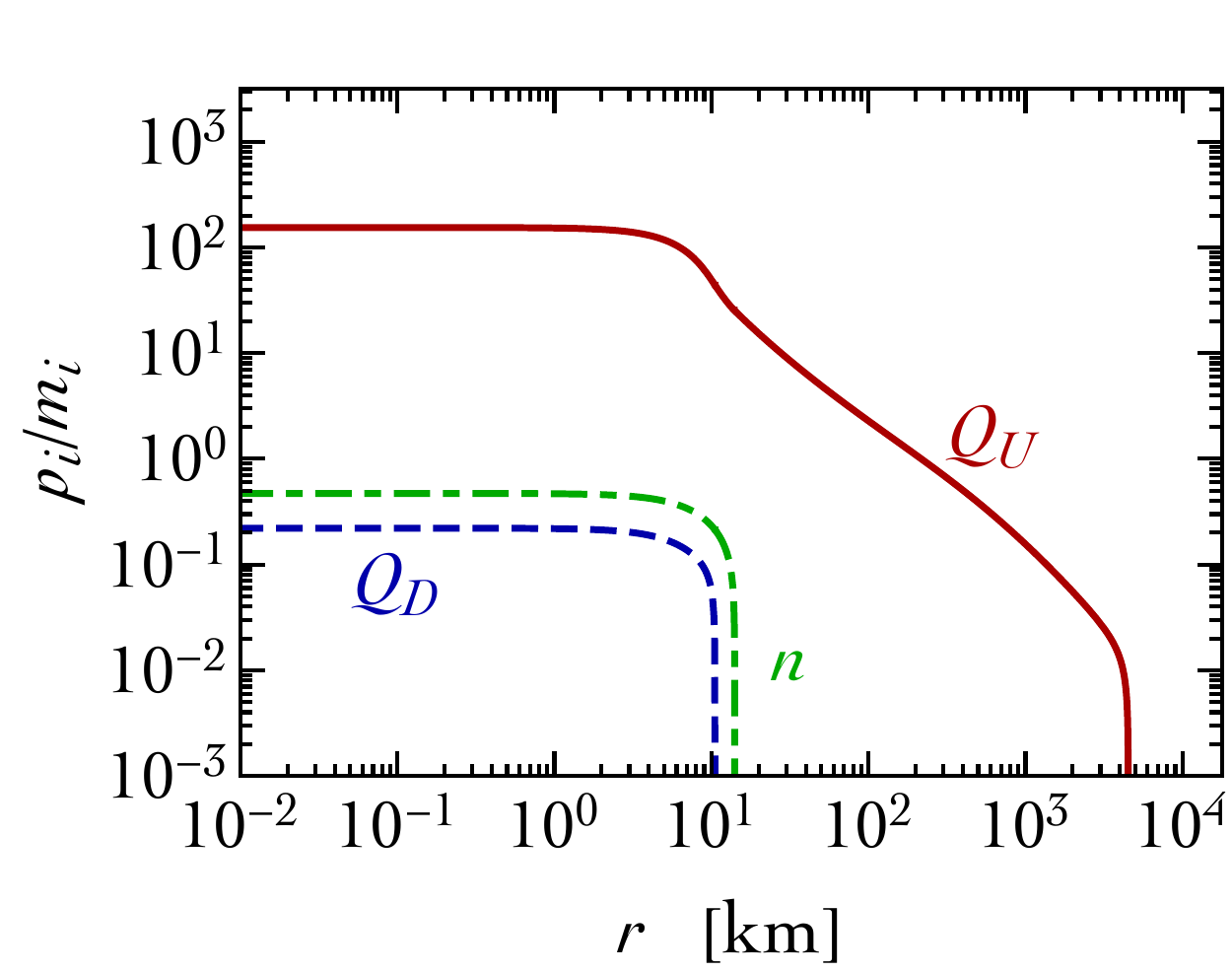}
\qquad\includegraphics[width=.35\textwidth]{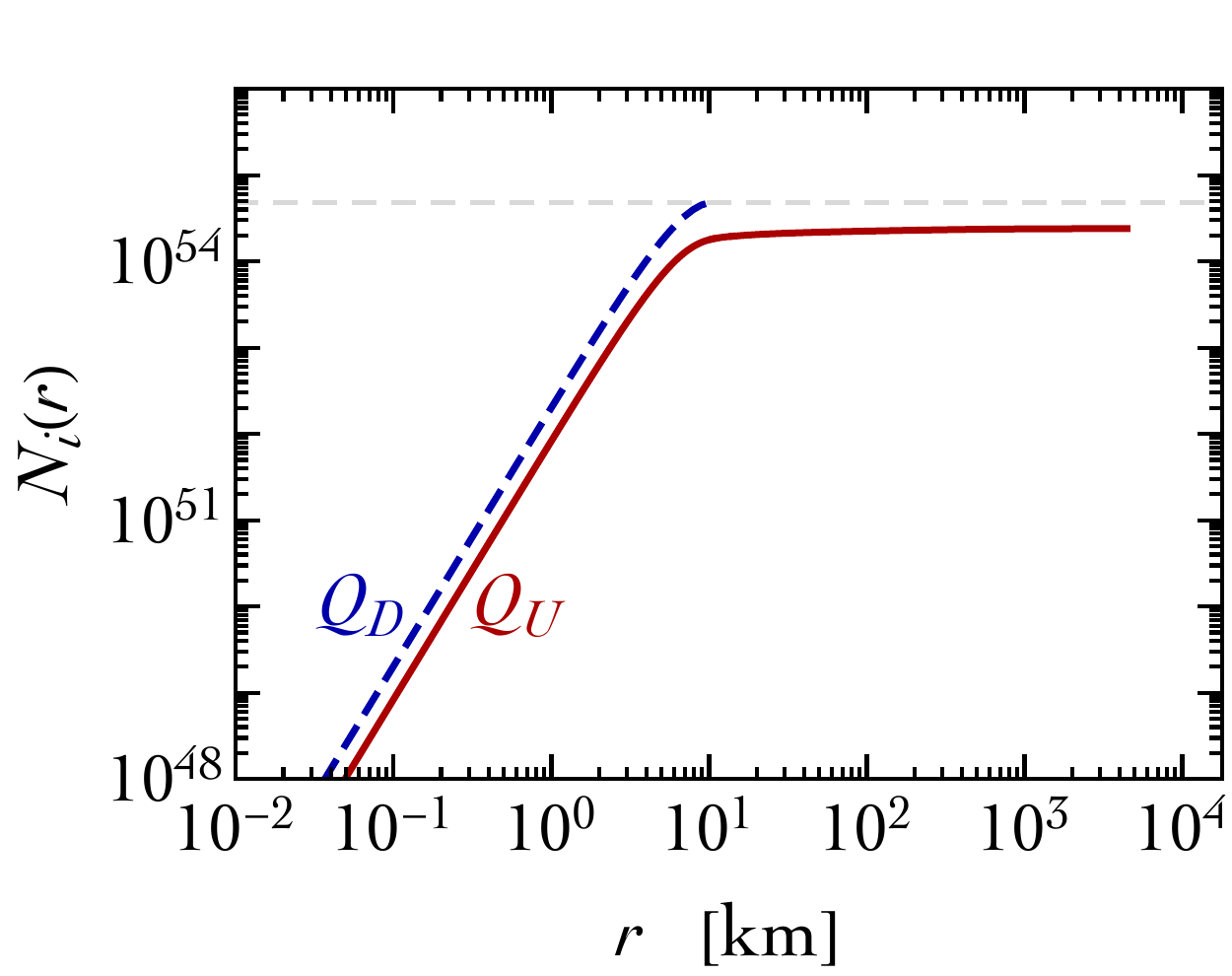}$$\\
\vspace{-1.cm}
$$\includegraphics[width=.35\textwidth]{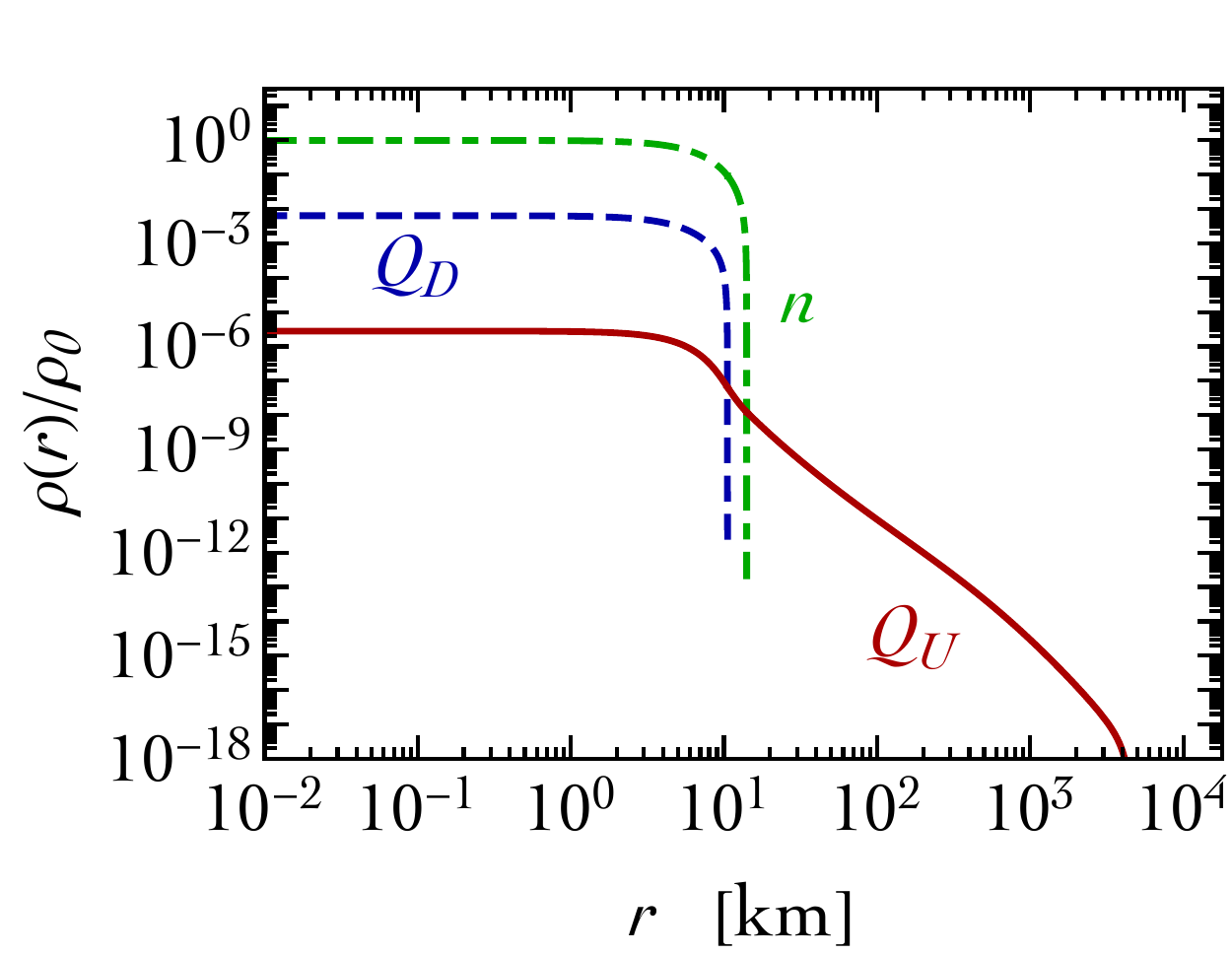}
\qquad\includegraphics[width=.35\textwidth]{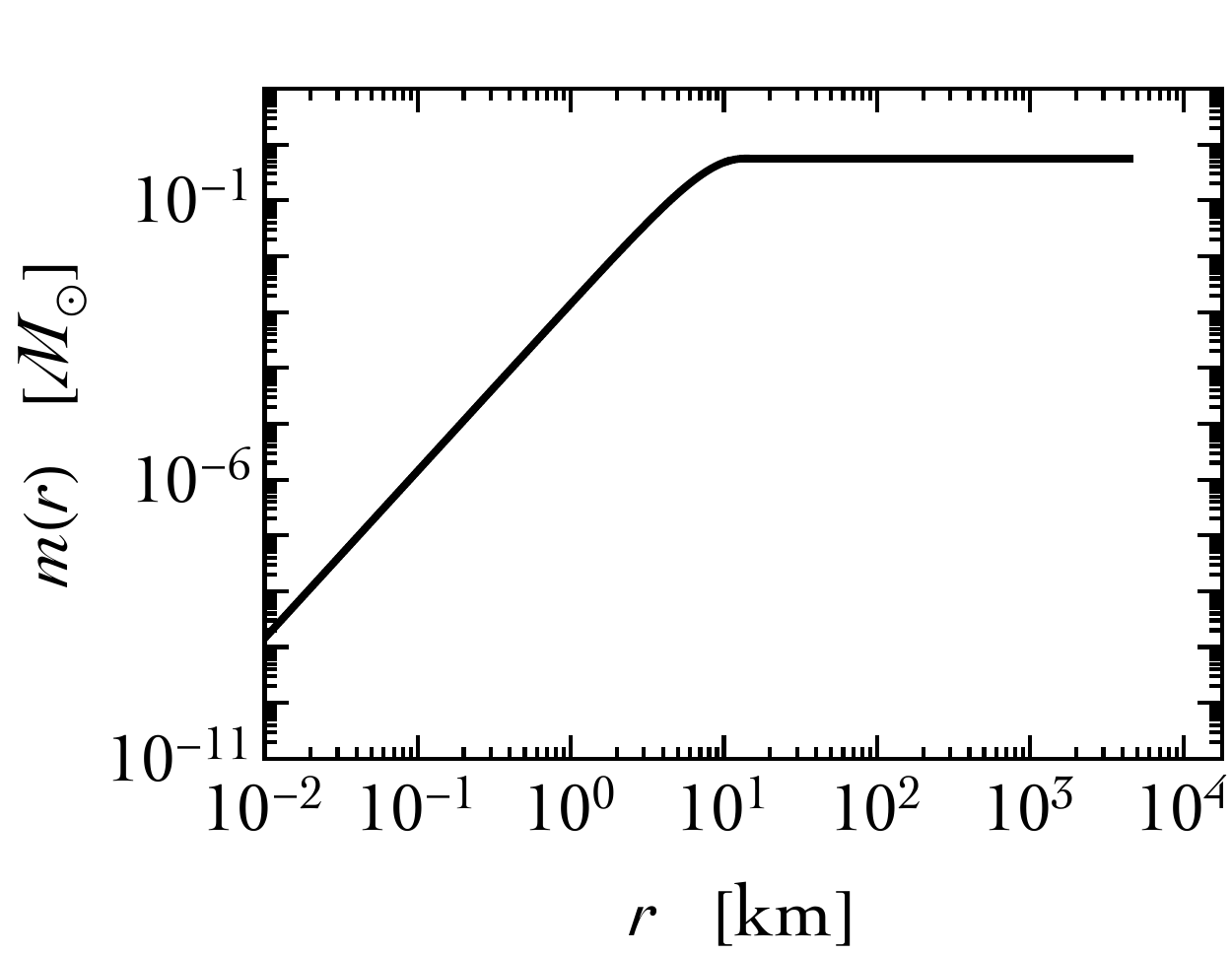}$$
\caption{\em \label{fig:ChargedStarNS} 
Same as in fig.\,\ref{fig:ChargedStar} but with the additional presence of neutrons (dot-dashed green) besides the two charged species $Q_D$ (dashed blue) and $Q_U$ (solid red).
 }
\end{center}
\end{figure}
The condition of charge neutrality is $2n_{U}(r) = n_{D}(r)$ (that is $\rho_e(r) = 0$). 
We stress that this condition, if imposed, is valid for any $r$.
 However, as done in section~\ref{sec:ChargedStar}, we shall not impose the condition of charge neutrality in our computation.
There is indeed no compelling reason to impose the condition $2n_{U}(r) = n_{D}(r)$  at any values of $r$.
On the contrary, we are more interested in computing the density distributions $n_i(r)$ by solving 
the Einstein-Maxwell field equations  
as a result of the equilibration between 
 gravitational and electromagnetic interactions, and without any additional restriction. 
We refer to appendix~\ref{app:Tech} for technical details.  
 Our results are similar to those already discussed in section~\ref{sec:ChargedStar} with the additional presence of neutrons besides the two charged distributions. 
 
For simplicity, we assume Fermi degeneracy for all of them. In the case of neutrons, this is not realistic but we shall address this point in the next section. 
We assume the mass hierarchy $m_D \gg m_U$, and we take as explicit values $m_D = 469$ MeV and $m_U = 0.5$ MeV (assuming the opposite hierarchy, $m_U \gg m_D$ does not change the qualitative results of this section, and it will be discussed in more detail in section~\ref{sec:RealisticEoS}). 
We show our results in fig.\,\ref{fig:ChargedStarNS}. 
Because of the mass hierarchy in the charged decay product, at the equilibrium we have a net separation of charge since $R_{U} \gg R_{D}$, as evident from the top-left panel of fig.\,\ref{fig:ChargedStarNS} in which we show the radial distributions of the Fermi momenta. The radii of the Fermi sphere in position space corresponds to $n_i(R_i) = 0$ for the three distributions.

We impose the condition of charge conservation on the total number of charged particles, that is $\mathcal{N}_{D} = 2\mathcal{N}_{U}$. This is shown in the top-right panel in which we plot the number of charged particles as a function of the radial distance from the center of the star. 
The radius of the star is given by $R=R_n$. This definition is motivated---in parallel with the discussion in section~\ref{sec:ChargedStar}---by the  mass density of the star receiving a negligible contribution from the region $r \geqslant R_n$ since the only contribution comes from the very light type-$U$ particles. This is illustrated in the bottom row of fig.\,\ref{fig:ChargedStarNS}. In the bottom-left panel we show the ratio $\rho(r)/\rho_0$ by separating the contribution of the three different species, where the total mass density of the star is defined by
\begin{equation}
\rho(r) = \frac{1}{3\pi^2}\left[
m_{n}p_n(r)^3 + m_{D}p_{D}(r)^3  + m_{U}p_{U}(r)^3 
\right]~.
\end{equation}
In the bottom-right panel of  fig.\,~\ref{fig:ChargedStarNS}, we show the mass-energy $m(r)$ of the star as a function of the radial distance. 
For $r\geqslant R_n$, it remains approximately constant thus justifying the definition of mass $M = m(R)$. 
We indicate with $M_Q$ the fraction of the total mass-energy density inside the radius $R$ that is due to the energy density of the charged particles and the energy density of the electric field.

\begin{figure}[!htb!] 
\begin{center}
	\includegraphics[width=.5\textwidth]{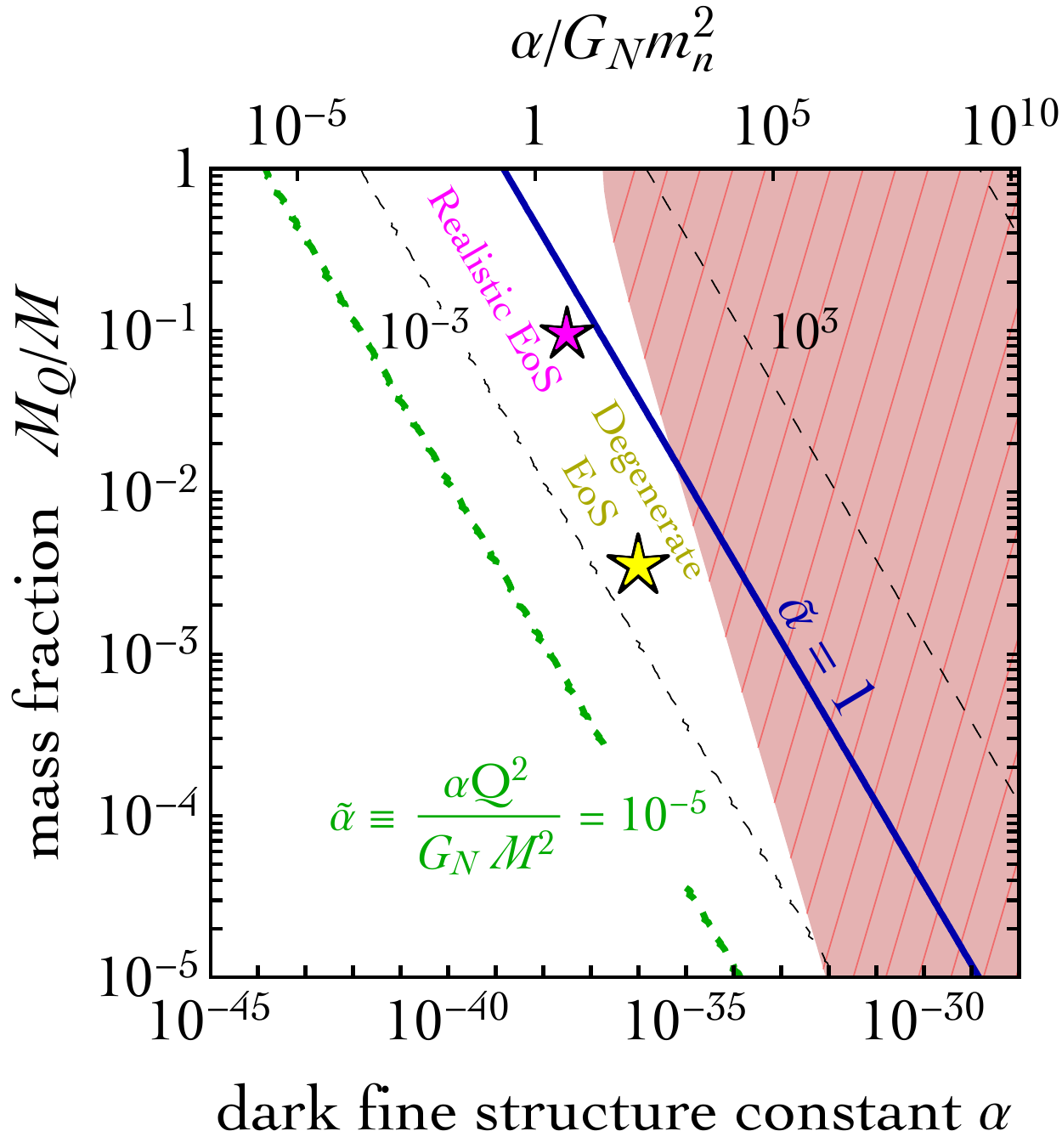}
	\caption{\em \label{fig:ChargedNS}   
	Same as in  fig.\,\ref{fig:StarPlot} but zoomed in the region that is relevant for the phenomenological analysis carried out in section~\ref{sec:GW}.
	The yellow star marks the explicit solution constructed in section~\ref{sec:ChargedNS}. 
	The magenta star marks an explicit realization of the realistic solutions found in section~\ref{sec:RealisticEoS}.
 }
\end{center}
\end{figure}

Finally, we can compute the electric charge.
Because of charge conservation, we have $Q(R_{U}) = 0$.
However, $Q(r)\neq 0$ if $r < R_{U}$ because of $2n_{D}(r) \neq n_{U}(r)$.
As a consequence of our definition of $R$, therefore, we have 
$\tilde{\mathcal{Q}} = Q(R)\neq 0$, and the outcome of the computation is a neutron star with a 
net electric charge.  The discussion about the Debye screening presented in section~\ref{sec:DarkStars} remains valid and applicable also in this case.

To discuss the role of these charged neutron stars in light of the phenomenological analysis outlined in section~\ref{sec:GW}, in fig.\,\ref{fig:ChargedNS} we zoom in the top-left corner of  fig.\,\ref{fig:StarPlot}.
We plot on the $x$-axis the value of the dark fine structure constant $\alpha$ and on the $y$-axis the dark fraction of the neutron star. The red region is excluded by the structural bound discussed in section~\ref{sec:DarkStars}.
The diagonal dashed lines correspond to contours of constant $\tilde{\alpha}$ with the value $\tilde{\alpha} = 10^{-5}$ that corresponds to a realistic lower limit (subject to the caveats discussed in section~\ref{sec:GW}) detectable with gravitational wave interferometers. 
In this plane, the yellow star marks the explicit solution constructed in this section obtained with $\alpha = 10^{-36}$. 
Testing the existence of such a charged star is well within the reach of the sensitivity of gravitational wave interferometers.

\subsection{Towards a realistic case}\label{sec:RealisticEoS}

The case studied in section~\ref{sec:ChargedNS} is not realistic because we assumed Fermi degeneracy for the neutrons.
As well known, the maximum mass of a neutron star that is supported against gravitational collapse only by degeneracy pressure is about $M_{\rm max}\sim 0.7\,M_{\odot}$, well below the maximum observed mass of a neutron star that is $M_{\rm max}^{\rm obs}= 2.01\pm 0.04\,M_{\odot}$ 
(measured  in the binary pulsar PSR J0348+0432, with 1-$\sigma$ error~\cite{Antoniadis:2013pzd}). This measurement sets an empirical lower bound on the maximum value of the  neutron star mass, and any attempt of modeling its equation of state must satisfy  it in order to be considered realistic. 

In addition, the observation of gravitational waves from the binary neutron star inspiral GW170817~\cite{TheLIGOScientific:2017qsa} puts a severe upper limit on the so-called dimensionless tidal deformability parameter $\Lambda$.
The latter characterizes the quadrupole deformation of a neutron star in response to an external gravitational field, and it strongly depends on the equation of state of the neutron star since, roughly speaking, a soft state of matter can be more easily deformed than a stiff one. 
The individual tidal deformability parameters of the two neutron stars, $\Lambda_1$ and $\Lambda_2$, cannot be disentangled in the observed gravitational waveform.  Instead, what is measured is an effective tidal deformability, dubbed $\tilde{\Lambda}$ in~\cite{TheLIGOScientific:2017qsa}, which is a mass-weighted average of $\Lambda_1$ and $\Lambda_2$.  The measurement of $\tilde{\Lambda}$ correlates with the spins of the two neutron stars.
The bound on $\tilde{\Lambda}$ in~\cite{TheLIGOScientific:2017qsa} is 
\begin{equation}
\tilde{\Lambda} \equiv \frac{16}{13}\left[
\frac{(M_1 + 12 M_2)M_1^4\Lambda_1 + (M_2 + 12 M_1)M_2^4\Lambda_2}{(M_1 + M_2)^5}
\right]  \leqslant 
\left\{
\begin{array}{cc}
800  &  90\%\,{\rm C.L.\,low\,spin\,priors} \\
 & \\
 700 &   \,\,90\%\,{\rm C.L.\,high\,spin\,priors} 
\end{array}\right.
\end{equation}

\begin{figure}[!htb!]
\begin{center}
$$\includegraphics[width=.48\textwidth]{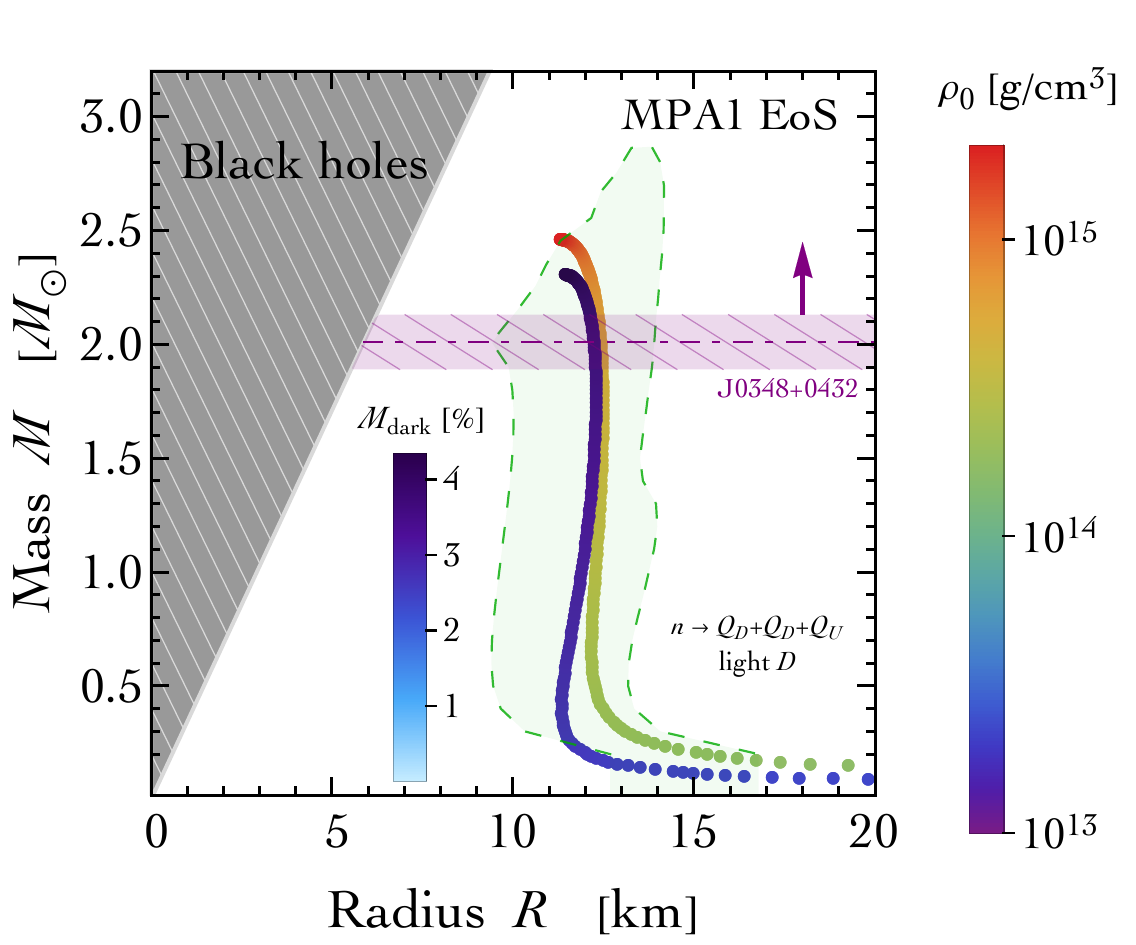}
\qquad\includegraphics[width=.48\textwidth]{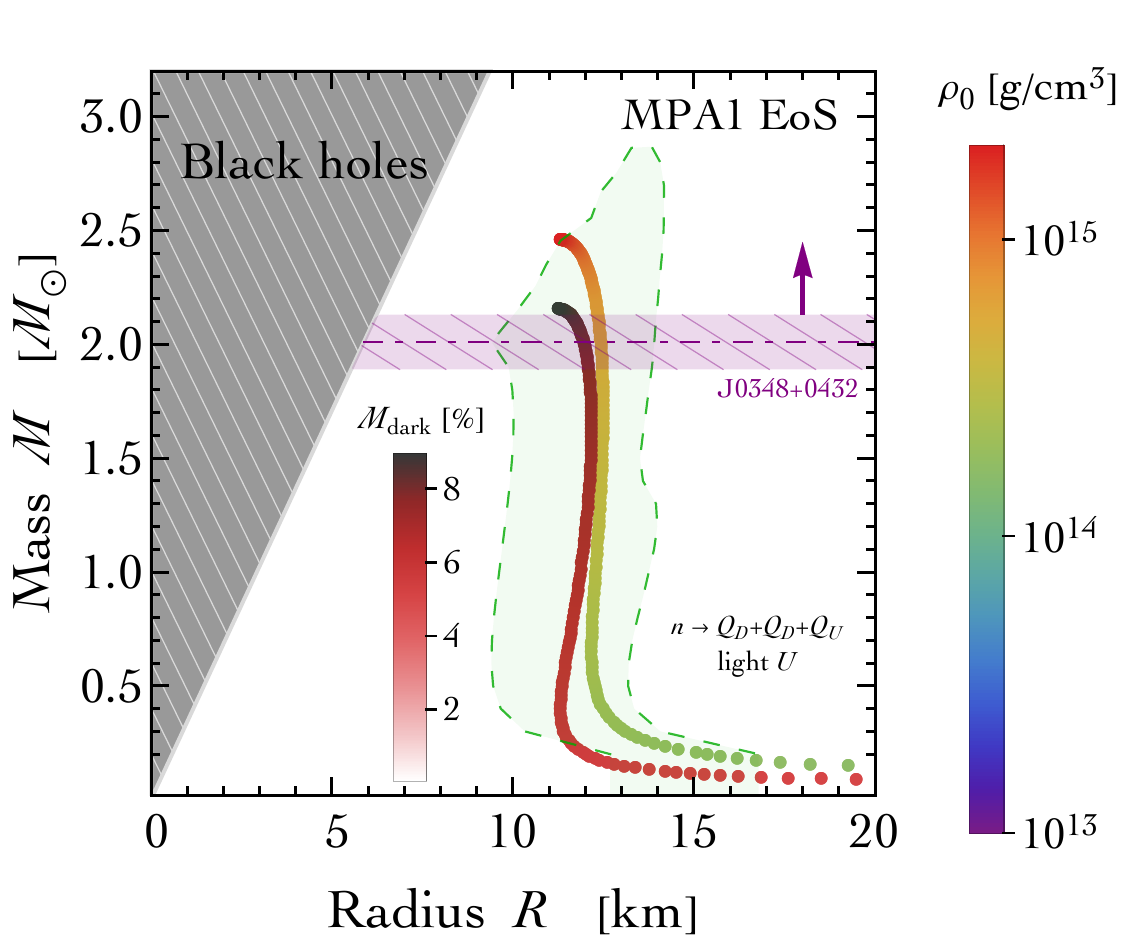}$$
\caption{\em \label{fig:SoftEoS} 
	Mass-radius relation for realistic neutron stars without (rainbow colors) and with (deep-blue colors and cherry tones) neutron decay.
 In the case of a pure neutron star, the legend with rainbow colors keeps track of different values of the central density. 
 When neutron decay is included, the inset legend in the left (right) plot with deep-blue colors (cherry tones) keeps track of the dark fraction in the neutron star mass assuming the neutron decay process $n \to Q_D + Q_D + Q_U$ with light $D$ (with light $U$).
	 The green region with dashed boundary (from~\cite{Annala:2017llu}) is compatible with the observations of neutron stars with mass $M \simeq 2 M_\odot$ and with the bound on the tidal deformability parameter $\Lambda(1.4\,M_{\odot}) < 800$~\cite{TheLIGOScientific:2017qsa}. 
	 The purple horizontal band corresponds to the 3-$\sigma$ error   
	in the measurement of the neutron star mass in the binary pulsar PSR J0348+0432. 
	The neutron star equations of state are taken from~\cite{EoS}. We used the MPA1 equation of state.
 }
\end{center}
\end{figure}

 By means of a linear expansion around the \textit{canonical} reference mass $M = 1.4\,M_{\odot}$~\cite{DelPozzo:2013ala}, 
it is possible to recast this result into a bound on the tidal deformability parameter $\Lambda$:  
$\Lambda(1.4\,M_{\odot}) \leqslant 800$ with the low-spin prior ($\Lambda(1.4\,M_{\odot}) \leqslant 1400$ with the high-spin prior) \cite{TheLIGOScientific:2017qsa}. 
The bound $\Lambda(1.4\,M_{\odot}) \leqslant 800$ was used in~\cite{Annala:2017llu} to extract a constraint on the neutron star equation of state.
In our analysis we use the results of~\cite{Annala:2017llu} (see also~\cite{Most:2018hfd}) that extracted a non-trivial constraints on a generic family of neutron star  equations of state that interpolate between state-of-the-art theoretical results at low and high baryon density. The corresponding allowed region in the mass-radius plane is shown in green in  fig.\,\ref{fig:SoftEoS}. A realistic neutron star equation of state must, therefore,
give rise to a mass-radius curve that fits in the green region, as well as be capable of sustaining a maximum mass at least equal to about $2\,M_{\odot}$. 
 There exists a number of proposed equations of state that satisfy these criteria, and we use the results tabulated in~\cite{EoS}.
 
 There is, however, an additional (and very important) problem.
As shown in~\cite{McKeen:2018xwc}, explaining the neutron lifetime puzzle by means of neutron decay into dark fermions is not compatible with the aforementioned properties of neutron stars. 
The reason is that the presence inside the neutron star of the dark decay products of the neutron, if described by a non-interacting degenerate Fermi gas, softens the neutron star equation of state so much that it is no longer possible to support a mass at or above  $2\,M_{\odot}$, thus in conflict with observations. The validity of this argument 
depends on the amount of dark particles that accumulates inside the neutron star as a consequence of neutron decay. This quantity, in turn, is fixed by the condition of chemical equilibrium that describes the equilibration of the neutron decay process with its inverse. 

In the following, we demonstrate  that the decay process in eq.~(\ref{eq:DarkDecay}) represents---thanks to the presence of a conserved charge---a simple exception to the applicability of  the argument proposed in~\cite{McKeen:2018xwc}.\footnote{Two possible ways out were previously proposed. In~\cite{Cline:2018ami}, the presence of repulsive self-interactions among the dark particles  produces a hardening of the equation of state. In~\cite{Grinstein:2018ptl},  repulsive  interactions  between  dark particles  and  neutrons make neutron conversion thermodynamically disfavored.
}
For the sake of clarity, we start  by first simplifying  the discussion as much as possible. 
In our subsequent numerical analysis, on the contrary, we  provide a full treatment in the context of general relativity. 

We consider baryonic matter made of neutrons with pressure $P_n$, energy density $\varepsilon_n$ and number density $n_n$. The neutron chemical potential is $\mu_n = (P_n+\varepsilon_n)/n_n$. 
 \begin{figure}[!htb!]
\begin{center}
	\includegraphics[width=.5\textwidth]{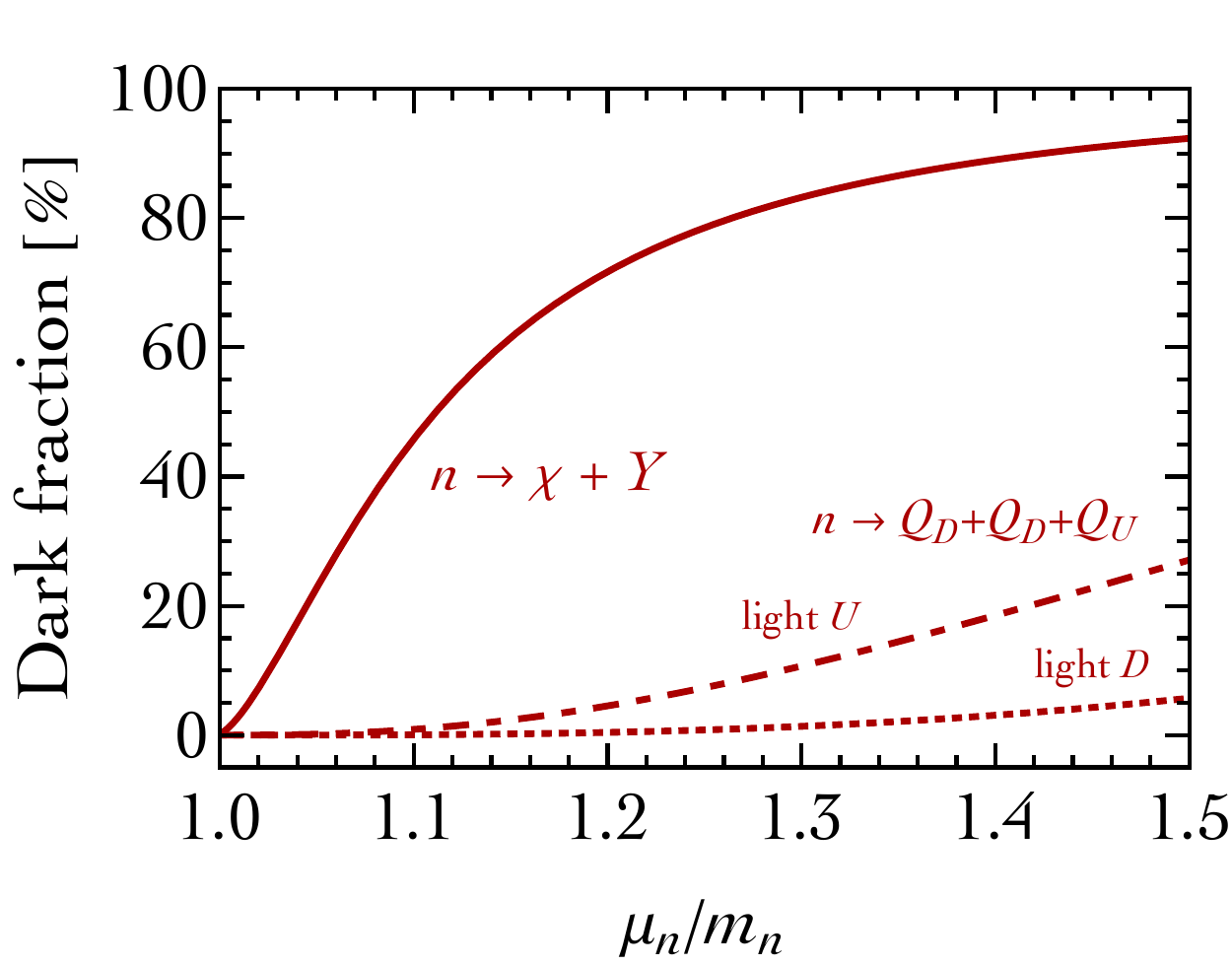}
	\caption{\em \label{fig:DarkFraction}  	
Sketch of the dark fraction of a neutron star with mass $M =1.5\,M_{\odot}$ and radius $R=10$ km as a function of the neutron chemical potential originated from two different neutron decay channel (see text for details). 
We show in solid red the case $n\to \chi + Y$ analyzed in~\cite{McKeen:2018xwc}. The other two lines refer to the decay 
$n \to  Q_D + Q_D + Q_U$ studied in~\cite{Barducci:2018rlx}. We consider two possibilities: 
$m_D \gg m_U$ (dot-dashed red line) and $m_U \gg m_D$ (dotted red line).
 }
\end{center}
\end{figure}
As far as the neutron decay process is concerned, we consider the decay mode $n \to  Q_D + Q_D + Q_U$ studied in section~\ref{sec:ChargedNS}. 
  Let us first assume the mass hierarchy $m_D \gg m_U$, and for our numerical analysis we take $m_D\simeq 469$ MeV and $m_U = 0.5$ MeV.
   We impose chemical equilibrium and charge neutrality\footnote{This is different with respect to what done in section~\ref{sec:ChargedNS}---where we imposed chemical equilibrium but not charge neutrality---and the reason is twofold. First, imposing charge neutrality simplifies the computation since it relates the Fermi momenta of the charged final state particles. Second, it permits to sketch out the argument in a general way that does not depend on $\alpha$. However, we checked that the result of this section remains valid also if one assumes chemical equilibrium and charge conservation instead of charge neutrality, in the spirit of section~\ref{sec:ChargedNS}.}  
   to find
    \begin{eqnarray}
{\rm Chemical\,equilibrium:}~~~~~ \mu_n &=& 2\mu_{D} +  \mu_{U} = 2\sqrt{p_{D}^2 + m_{D}^2} + \sqrt{p_{U}^2 + m_{U}^2}~,\label{eq:ChemicalEqDDU} \\
{\rm Charge\,neutrality:}\,~~~~~~~~~   n_{D} &=& 2n_{U}~.\label{eq:ChargeneutralityDDU}
    \end{eqnarray}
    These two equations fix the values of the Fermi momenta. Eq.~(\ref{eq:ChargeneutralityDDU}) gives $p_U = p_D/2^{1/3}$.
    Even though the corresponding expression for $p_D$ that follows from eq.~(\ref{eq:ChemicalEqDDU}) is quite lengthly, we can use the illustrative limit $m_U \to 0$ and $\mu_n \gg m_D$. We find $p_D \simeq 0.4\,\mu_n + O(m_D^2/\mu_n)$. 
    The fraction of $\mu_n$ that is missing is carried away by the light type-$U$ dark particle. 
    
     It is now possible to compute the number density of the heavy type-$D$ dark particle since $n_{D} = p_{D}^2/3\pi^2$. This is important because we are interested in the dark fraction of the neutron star at the equilibrium. For a crude estimate, we define the relative abundance $\Delta n_{\rm dark} = n_{D}/(n_{D} + n_{\rm NS})$, where 
  $n_{\rm NS}\equiv 3(M/m_n)/4\pi R^3$ with $M = 1.5\,M_{\odot}$ ad $R= 10$ km.
 
  We show the value of $\Delta n_{\rm dark}$ in  fig.\,\ref{fig:DarkFraction} (dot-dashed red line). For realistic values of $\mu_n$ (for a free neutron $\mu_{n} = m_n$ while in a neutron star for realistic equations of state we have $\mu_{n} \gtrsim m_n$) we expect a dark fraction that is much smaller if compared with the case in which one considers the decay $n\to \chi + Y$. This is the case studied in~\cite{McKeen:2018xwc}, in which $\chi$ is a single neutral dark Dirac fermion and $Y$ is a possibly multi-particle final state with zero net chemical potential. Following the same logic outlined before, one obtains in this case dark fraction of order $60\%$ for realistic values of $\mu_n$ (fig.\,\ref{fig:DarkFraction}, solid red line).
  
  The decay $n \to Q_D + Q_D + Q_U$ with the mass hierarchy $m_U \gg m_D$---opposite to the one just studied above---should give the optimal situation since a large fraction of the neutron chemical potential will be carried away by the light type-$D$ dark particles.
  Being light, they will not contribute much to the mass fraction of the neutron star even with a large Fermi sphere. Working in the limit $m_D \to 0$ and $\mu_n \gg m_U$, we find $p_U \simeq 0.3\,\mu_n + O(m_U^2/\mu_n)$. 
  We show the value of $\Delta n_{\rm dark}$ in  fig.\,\ref{fig:DarkFraction} (dotted red line). 
  We use $m_U = 938$ MeV and $m_D = 0.5$ MeV.
  For realistic values of $\mu_n$, we expect a negligible dark fraction.
 
 From this very simple discussion  it seems possible to evade the bound of \cite{McKeen:2018xwc} in the presence of a multi-body neutron decay into the dark sector. In order to confirm our estimate, we solve numerically the Einstein field equations for the cases discussed above.
We refer to appendix~\ref{app:Tech} for technical details. 
We consider the neutron star equation of state tabulated in~\cite{EoS}, and we focus in particular on the MPA1 equation of state as a starting point since it agrees well---in the absence of any dark component---with the bound on the tidal deformability parameter and the maximum neutron star mass. 

In  fig.\,\ref{fig:SoftEoS} the mass-radius relation obtained for such equation of state in the absence of a dark component is shown with rainbow colors, whose gradation illustrates the different values of the baryonic density 
at the center of the star. In  the left and right panel of  fig.\,\ref{fig:SoftEoS} we show the impact on the mass-radius relation of a dark component originating from neutron decay $n \to Q_D + Q_D + Q_U$. 

In the left (right) panel we consider the case in which the type-$D$ (type-$U$) final state particle is light.
In both cases, the intuition of the previous discussion and the na\"{\i}ve result of fig.~\ref{fig:DarkFraction} is confirmed: The dark fraction inside the neutron star does not exceed few percent, and the mass-radius relation predicted by the realistic MPA1 equation of state remains consistent with the bounds on tidal deformability and maximum mass.

For completeness, we mark with a magenta star in  fig.\,\ref{fig:ChargedNS} a typical solution belonging to those in the right panel of  fig.\,\ref{fig:SoftEoS}. By relaxing the requirement of charge neutrality in favor of the more general condition of charge separation, the solution lies well within the reach of gravitational-wave interferometers.

\section{Discussion}

It is natural to frame the range of sensitivity of gravitational-wave detection in terms of searches of possible deviations from (Newtonian) inverse-squared law gravity (for limits, see, for example, \cite{Adelberger:2003zx}). The latter are usually parametrized in terms of the  potential
\be 
V(r) = - \frac{G_N M_1 M_2}{r} \left[ 1 - \tilde{\alpha} e^{-r/\lambda} \right]~, \label{gravity}
\ee
where the Yukawa-like term is often dubbed the fifth force. The potential in \eq{gravity}
 is easily mapped into that in eq.~(\ref{eq:DarkPotDeb}) in section~\ref{sec:DarkDipole}.

 \begin{figure}[!htb!]
\begin{center}
	\includegraphics[width=.5\textwidth]{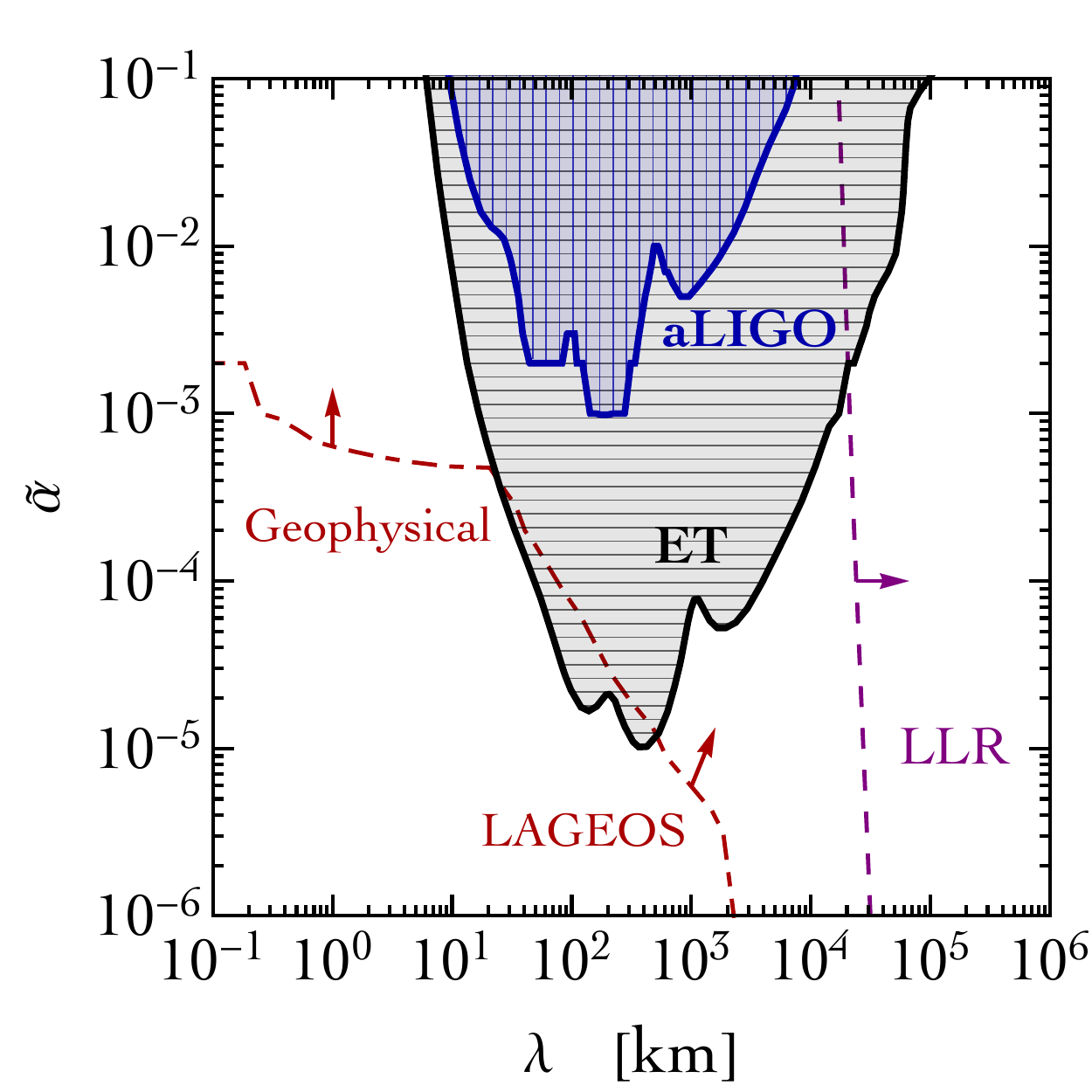}
	\caption{\em \label{fig:ISLGravity}  	
Gravitational-wave sensitivity (aLIGO in its current operative set-up and a prospect for the Einstein Telescope, ET) to dark forces and current limits on deviations from Newton gravity from geophysical, satellite (LAGEOS) and  lunar-laser-ranging (LLR) observations (direction of the arrows excluded). 
The interpretation of this plot is twofold. 
If one assumes a universal correction to gravitational interactions in the form of an additional Yukawa contribution as in eq.~(\ref{gravity}), geophysical, LAGEOS and LLR limits  apply. In this case the length scale on the horizontal axis corresponds to the finite range of the Yukawa potential due to the exchange of virtual-force carriers of mass $1/\lambda$. If one considers the situation studied in this paper in which a U(1) dark force  manifests itself at the macroscopic level only for interactions between neutron stars as the consequence of their net dark charge, geophysical, LAGEOS and LLR limits do not apply.
In this case, distances on the horizontal axis are measured in terms of the Debye screening which plays in our approach the role of the $\lambda$ parameter in eq.~(\ref{gravity}).
 }
\end{center}
\end{figure}

We use the results of \cite{Alexander:2018qzg} to plot the current (aLIGO) and expected (ET) sensitivity of gravitation-wave detection experiments and compare with those known from geophysical, LAGEOS satellites  and lunar-laser-ranging (LLR)  observations in fig.\,\ref{fig:ISLGravity}. 

The comparison shows the geophysical, LAGEOS and LLR  limits performing better at most of the distances probed. Only around distances of the order 100 km, the expected limits from the Einstein telescope (ET)~\cite{Hild:2009ns} become competitive and even better. That said, we must bear in mind that  such a comparison brings together very different physical settings. Limits and detection prospects from gravitational-wave experiments deal with a dark force that manifests itself at the macroscopic level only in the interaction between objects that are charged (neutron stars in our case) because of the presence of  dark sector states and  at distances  below those for which   Debye screening  takes place. Whereas such a force cannot be probed by any geophysical or laser-ranging experiment because they only test interactions for which ordinary matter is charged---gravitational-wave astronomy, as we argued,  might be sensitive  to the dark sector if it contains a new gauge interaction  weaker than gravity.



\appendix

\section{Charged stars in Einstein-Maxwell theory}\label{app:Tech}

In this appendix we discuss the Einstein-Maxwell field equations describing the equilibrium configuration of a charged star.
We use the mostly-plus flat Minkowski metric $\eta = (-1,+1,+1,+1)$.
We start from the Einstein-Maxwell field equations in the presence of charged matter. We have
\begin{eqnarray}\label{eq:EE}
R_{\mu\nu} - \frac{1}{2}g_{\mu\nu}R &=& 8\pi G_N T_{\mu\nu}~,\label{eq:EE}\\
\nabla_{\nu}F^{\mu\nu} &=& J^{\mu}~,\label{eq:ME} 
\end{eqnarray}
where $R$ is the Ricci scalar and 
$F_{\mu\nu} \equiv \nabla_{\mu}A_{\nu} - \nabla_{\nu}A_{\mu} = \partial_{\mu}A_{\nu} - \partial_{\nu}A_{\mu}$
 the electromagnetic field strength.
We use the notation of electromagnetism but we refer to a generic dark $U(1)$ gauge symmetry.
We use natural units with $c=\hbar=1$ and $\epsilon_0 = 1$ for the vacuum permittivity (so that for the fine structure constant we have $\alpha = e^2/4\pi$).
We adopt the following stationary spherically symmetric ansatz for the metric\footnote{Spherically-symmetric static configurations are idealized objects.  
The unavoidable  presence  of  rotation  breaks  spherical  symmetry  giving  rise  to  a  configuration  that  is axisymmetric. Throughout this paper, we neglect for simplicity the role of rotation. 
}
 \begin{equation}\label{eq:Metric}
 ds^2 = g_{\mu\nu}dx^{\mu}dx^{\nu} =  -B(r)dt^2 + A(r)dr^2 + r^2\left(
 d\theta^2 + \sin^2\theta d\phi^2
 \right)~,
 \end{equation}
with $x^{\mu} = (t,r,\theta,\phi)$ the conventional Schwarzschild-like coordinates. 
Notice that, given the line element in eq.~(\ref{eq:Metric}), the spatial infinitesimal volume 
element 
 has the measure  $\sqrt{g_{ij}} = \sqrt{A}r^2\sin\theta$. 
We separate the energy-momentum tensor
 in two contributions, namely $T_{\mu\nu} \equiv M_{\mu\nu} + E_{\mu\nu}$.
 In this decomposition, $M_{\mu\nu}$ represents the  energy-momentum tensor of a perfect fluid 
  with mass-energy density $\varepsilon(r)$  and  pressure $P(r)$ 
while $E_{\mu\nu}$ is the electromagnetic energy-momentum tensor
 \begin{equation}
 M_{\mu\nu} = \left(
\begin{array}{cc}
 \varepsilon(r) B(r)  & 0   \\
 0 & P(r) g_{ij}    
\end{array}
\right)~,~~~~~E_{\mu\nu} = \left[
 F_{\mu}^{\,\,\rho}(r)F_{\nu\rho}(r) - \frac{1}{4}g_{\mu\nu}F_{\rho\sigma}(r)F^{\rho\sigma}(r)
 \right]~.
 \end{equation}
 The  electric current density $J^{\mu}$ is
\begin{equation}
J^{\mu} = \rho_e u^{\mu} =\left(
\begin{array}{cc}
 \rho_e(r)/\sqrt{B(r)}  & 0   \\
 0 & 0    
\end{array}
\right)~,
\end{equation}
where $\rho_e(r)$ is the electric charge density. 
If we look for a static, spherically symmetric solution, the only non-zero components 
of the field strength are $F^{tr}(r) = -F^{rt}(r)\neq 0$. 
The time component of the Maxwell field equation, eq.~(\ref{eq:ME}), is
\begin{equation}\label{eq:Master1}
\frac{dQ(r)}{dr} = 4\pi r^2\sqrt{A(r)}\rho_e(r)~,
\end{equation}
where $dQ(r)$ is the electric charge in the infinitesimal shell 
between $r$ and $r+dr$.
We can obtain the charge inside a sphere of radial dimension $r$ if we integrate eq.~(\ref{eq:Master1}). 
The total charge of the system is given by
\begin{equation}\label{eq:TotalCharge}
Q(R) \equiv \tilde{\mathcal{Q}} =  4\pi\int_0^{R}d\tilde{r}\tilde{r}^2\rho_e(\tilde{r}) \sqrt{A(\tilde{r})}~,
\end{equation} 
where $R$ is the radius of the star.
The electric field that is given by
\begin{equation}\label{eq:ElectricField}
E(r) = \frac{Q(r)}{4\pi r^2} =  \frac{1}{4\pi r^2}\left[
4\pi\int_0^{r}d\tilde{r}\tilde{r}^2\rho_e(\tilde{r}) \sqrt{A(\tilde{r})}
\right]~.
\end{equation} 
In the numerical analysis, we need a differential form for eq.~(\ref{eq:ElectricField}). We find
\begin{equation}\label{eq:ElectricField2}
\frac{dE(r)}{dr} = -\frac{2E(r)}{r} + \rho_e(r)\sqrt{A(r)}~.
\end{equation} 
We now move to consider the $tt$-component of the Einstein field equations in eq.~(\ref{eq:EE}).
We find
\begin{equation}\label{eq:Aequation}
\frac{d}{dr}\left(
\frac{r}{A}
\right) = 1 - 8\pi G_N r^2\left(
\varepsilon + \frac{Q^2}{32\pi^2 r^4}
\right) = 1 - 8\pi G_N r^2\left(
\varepsilon + \frac{E^2}{2}
\right)~.
\end{equation}
This equation can be integrated. 
The strategy is to define  a new quantity $m(r)$ representing the
mass-energy inside the shell of radial coordinate $r$ in such a way that 
\begin{equation}
\label{eq:Master2}
 A(r) \equiv \left[
 1 - \frac{2G_N m(r)}{r}
 \right]^{-1}~,~~~~~~\frac{dA(r)}{dr} = 8\pi G_N rA(r)^2\left[
\varepsilon(r) + \frac{1}{2} E(r)^2
\right] - \frac{A(r)^2}{r}\left[
1 - \frac{1}{A(r)}
\right]~.
\end{equation}
Using eq.~(\ref{eq:Aequation}), we find
\begin{equation}\label{eq:MassEnergy}
 \frac{dm(r)}{dr} = 
4\pi r^2\left[\varepsilon(r) + \frac{1}{2} E(r)^2\right]~.
\end{equation}
The mass of the star is due to the total contribution of the energy density of the matter and the 
electric energy density. The mass-energy enclosed at radial distance $r$ takes the form
\begin{equation}\label{eq:ChargedMass}
m(r) = 4\pi\int_0^{r}d\tilde{r}\tilde{r}^2
\left[\varepsilon(\tilde{r}) + \frac{1}{2} E(\tilde{r})^2\right]~.
\end{equation}
We indicate with $M = m(R)$ the mass of the star. We now move to consider the spatial components of the Einstein field equations. 
Because of spherical symmetry, we can focus on the $rr$ component. The $\theta\theta$ and $\phi\phi$ components do not 
add any additional informations.
We find
\begin{equation}
\label{eq:Master3}
\frac{1}{B(r)}\frac{dB(r)}{dr} = 
\frac{1}{r}\left[
A(r) - 1
\right] + 8\pi G_N A(r) r\left[
P(r) - \frac{1}{2}E(r)^2
\right]~.
\end{equation}
Finally, we impose the Bianchi identity $\nabla_{\mu}T^{\mu\nu} = 0$.
The only non-trivial 
equation comes from the $\nu = r$ component. We find
\begin{equation}
\left(
P + \varepsilon
\right)B^{\prime} = \frac{B Q Q^{\prime}}{8\pi^2 r^4} - 2BP^{\prime}~.
\end{equation}
We rewrite $Q^{\prime}$ by means of the Maxwell field equation in eq.~(\ref{eq:Master1}), and, 
solving for $P^{\prime}$, we find
\begin{equation}\label{eq:TOV}
P^{\prime} = - \frac{(P + \varepsilon)}{2B}B^{\prime}
+ \rho_eE\sqrt{A}~.
\end{equation}
The chemical potential $\mu$ is defined by $P + \varepsilon = \mu\,n$, where $n$ is the number density.
It is  possible to write (with the help of eq.~(\ref{eq:Master3}) and using the definition of $A$) eq.~(\ref{eq:TOV}) in the form
\begin{equation}\label{eq:ChargedTOV}
\frac{dP(r)}{dr} = -\frac{G_N[P(r) + \varepsilon(r)]}{r\left[r - 2G_N m(r)\right]}\left\{
m(r) + 4\pi r^3\left[
P(r) - \frac{1}{2}E(r)^2
\right]
\right\} + \rho_e(r)E(r)\sqrt{A(r)}~,
\end{equation}
that is the generalization of the Tolman-Oppenheimer-Volkoff  equation in the presence of charge. In the absence of the electric field, $E(r)\to 0$, the gradient pressure is negative.
In this case the pressure starts from some high value at the center of the star, and it 
decreases going outwards. The radius of the star can be defined by the condition $P(R)=0$. In the presence of an electric field, the situation changes.
The main effect is the presence of the electrostatic pressure $P_E(r) \equiv  E(r)^2/2$.
The electrostatic term has always an opposite sign compared to gravity
 because the attraction due to gravity is contrasted by the Coulombian repulsion. 
By increasing the charge density, the 
electrostatic pressure  grows up to the point at which it overcomes
the gravitational attraction. The sign of the pressure gradient becomes positive, and it is no longer possible to find an equilibrium solution.  
This is the origin of the limit discussed in section~\ref{sec:DarkStars}.
 The Newtonian limit of eq.~(\ref{eq:ChargedTOV}) is
 \begin{equation}\label{eq:ChargedTOVNNew}
\frac{dP(r)}{dr} = -\frac{G_N\rho(r)}{r^2}\left[
m(r) 
- 2\pi r^3E(r)^2
\right] + \rho_e(r)E(r)~,
\end{equation}
where $\rho(r)$ is the mass density. Eq.~(\ref{eq:ChargedTOV}) describes hydrostatic equilibrium for charged matter in general relativity.
The system of differential equation that must be solved is formed by eqs.~(\ref{eq:ElectricField2},\ref{eq:MassEnergy},\ref{eq:Master2},\ref{eq:ChargedTOV}) with initial conditions $A(0) = 1$, $E(0) = 0$, $m(0) = 0$, $P(0)\equiv P_0$.
In order to close the system, an equation of state relating the pressure and the mass-energy density is needed.

In the presence of multiple species, the relation $P + \varepsilon = \mu n$ can be generalized 
\begin{equation}
P = \sum_{i}n_i\mu_i - \sum_i\varepsilon_i = \sum_i\left(
n_i \mu_i - \varepsilon_i
\right)~~\Longrightarrow~~ P_i = n_i \mu_i - \varepsilon_i~.
\end{equation}
This is the case of the two explicit constructions discussed in section~\ref{sec:ChargedStar} and section~\ref{sec:ChargedNS}. The electric charge density in this case takes the form 
\begin{equation}
\rho_e(r) = e\bigg[
\sum_{i}q_i n_i(r) - \sum_{j}q_j n_j(r)
\bigg]
\end{equation}
where we sum over the positively and negatively charged species with charges $+q_i e$ and $-q_j e$, respectively.
If charge neutrality is imposed, we have $\rho_e(r) = 0$ and a constraint among the number densities of the charged particles. In this case the electric field vanishes, and eq.~(\ref{eq:ChargedTOV}) reduces to the usual  Tolman-Oppenheimer-Volkoff equation.

Under the assumption of Fermi degeneracy, we have explicit relations for various microscopic quantities.
For clarity, we add a sub-index $_F$ to indicate that these relations are strictly valid for a degenerate Fermi gas (however, we omit this notation in the main text). 
The number density is related to the Fermi momentum $p_F$ by $n = p_F^3/3\pi^2$. 
The mass density is $\rho = mn$.
The Fermi energy $E_F$ is $E_F = (p_F^2 + m^2)^{1/2}$.
The pressure exerted by the degenerate fermions 
 can be computed by means of the mean momentum flux of the fermions. We find
\begin{equation}\label{eq:FermiPressure}
P = 
\frac{1}{\pi^2}\int_0^{p_F}dp\frac{p^4}{\sqrt{p^2 + m^2}}
=\frac{m^4}{8\pi^2}\left\{
x(1+x^2)^{1/2}(2x^2/3 - 1) + \log\left[
x + (1+x^2)^{1/2}
\right]
\right\}~,~~~~ x\equiv \frac{p_F}{m}~.
\end{equation}
The mass-energy density of the free fermions is related to the Fermi momentum by means of 
\begin{equation}\label{eq:EnergyDensityFermi}
\varepsilon = 
\frac{1}{\pi^2}\int_0^{p_F}dp\,p^2\sqrt{p^2 + m^2}
=
\frac{m^4}{8\pi^2}\left\{
x(1+x^2)^{1/2}(1 + 2x^2) - \log\left[
x + (1+x^2)^{1/2}
\right]
\right\}~.
\end{equation}
For the chemical potential, we find
\begin{equation}\label{eq:Ch}
\mu = m\sqrt{x^2 + 1}~.
\end{equation}
As expected, the chemical potential is the energy of the most energetic particle in
a degenerate fermi system, $E_F$.  
In general we have an implicit relation  between
 pressure $P$ and energy density $\varepsilon$. The only exception is the non-relativistic (NR) limit, 
 $x\ll 1$. We have $P_{\rm NR} = m^4x^5/15\pi^2$ and 
$\varepsilon_{\rm NR} = m^4x^3/3\pi^2$ (which coincides with the mass density $\rho$). 
In this limit, therefore, we find the polytrope equation of state 
$P_{\rm NR} = K\rho^{\gamma}$ with $\gamma = 5/3$. 
In section~\ref{sec:RealisticEoS} we used for the neutrons the tabulated equations of state collected in ref.~\cite{EoS}.

\section{On the role of post-Newtonian corrections}\label{app:PN}

In this appendix we discuss the impact of post-Newtonian corrections.

In section~\ref{sec:GW} we compared the leading terms of pure gravity and dark electromagnetism.  
The crucial point is that the dark dipole radiation introduces a functional dependence in the evolution of the orbital frequency $\omega$ 
that is different w.r.t. the gravitational quadrupole emission.
It is important to understand whether the inclusion of post-Newtonian gravitational corrections  
 can mimic the effect of the electromagnetic dipole term. If this happens, the constraining power of the analysis would  be limited by 
 an underlying  degeneracy between gravity and the additional dark radiation. We follow ref.~\cite{Blanchet:2013haa}, and we work at the 3.5\,PN order (however, we shall not include for simplicity the corrections due to the spin of each of the two stars in the binary).
 \begin{figure}[!htb!]
\begin{center}
	\includegraphics[width=.35\textwidth]{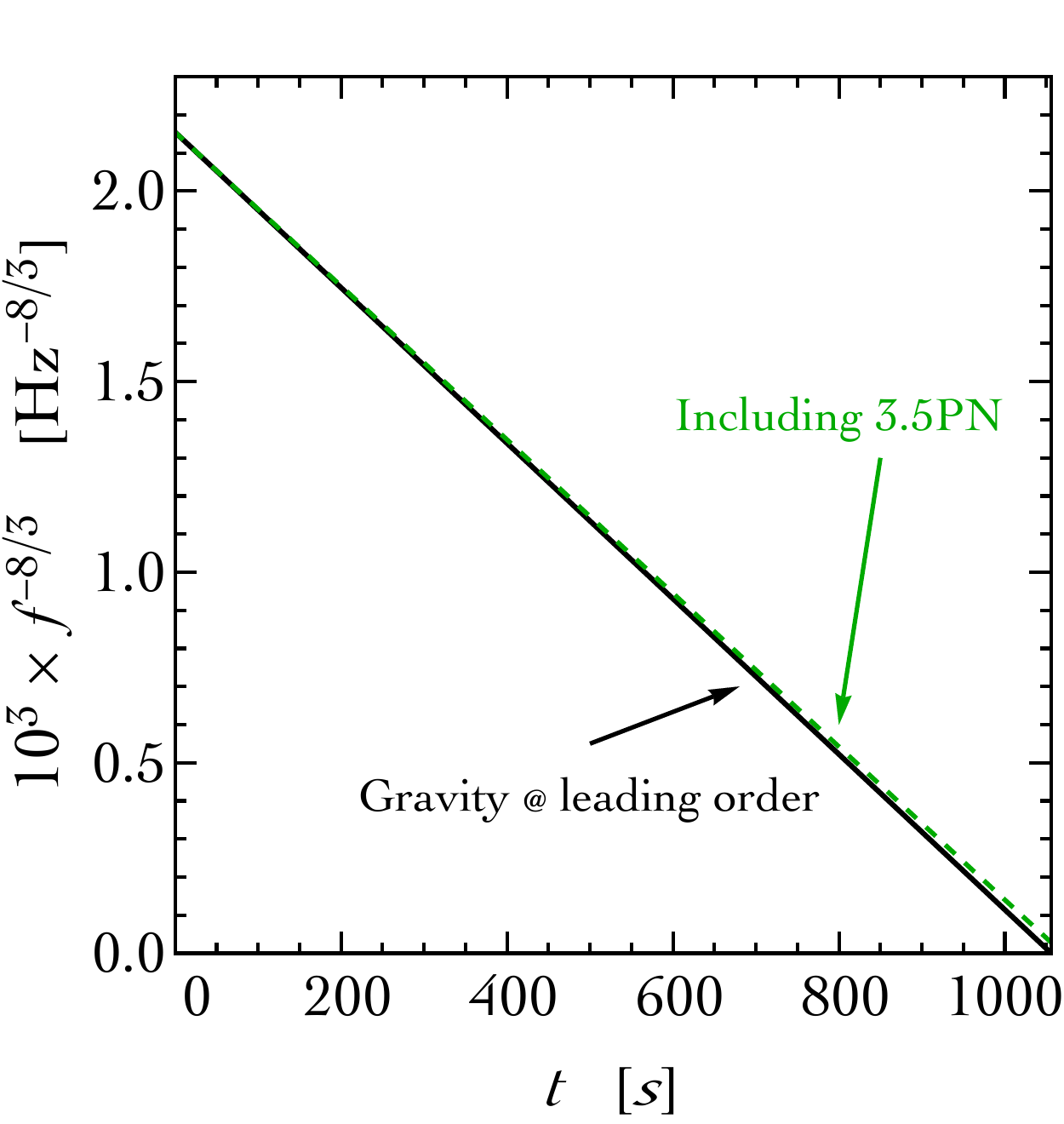}
	\caption{\em \label{fig:PNDyn} 
	Comparison between the time-evolution of the orbital frequency of the  inspiral using the leading order result in eq.~(\ref{eq:PureGravity}) (black solid line) and including  3.5PN corrections in
	eq.s~(\ref{Ecirc},\ref{fluxEx}) (dashed green line).
 }
\end{center}
\end{figure}
 The crucial equations are
the modified expression for the total energy of the binary system, $E_{\rm tot}$, 
and the radiated power in the form of gravitational waves beyond the Einstein's quadrupole formula, $\mathcal{P}_{\rm GW}$.
We express both these quantities as expansions in the parameter 
\begin{equation}
x\equiv \left[\frac{G_N(M_1 + M_2)\omega}{c^2}
\right]^{2/3} = O\left(\frac{1}{c^2}\right)~,
\end{equation}
where we introduced explicitly the speed of light $c$ to make the PN expansion more transparent. 
For the total energy, we use the expression~\cite{Blanchet:2013haa}
\begin{align}
  E_{\rm tot}^{\rm 3.5\,PN} &= -\frac{\mu c^2 x}{2} \biggl\{ 1 +\left(-\frac{3}{4} -
  \frac{1}{12}\nu\right) x + \left(-\frac{27}{8} + \frac{19}{8}\nu
  -\frac{1}{24}\nu^2\right) x^2 \nonumber \\ & \qquad \quad + \left[
  -\frac{675}{64} + \left(\frac{34445}{576} - \frac{205}{96}\pi^2
    \right)\nu - \frac{155}{96}\nu^2 - \frac{35}{5184}\nu^3 \right]
  x^3 \biggr\} + O\left(\frac{1}{c^8}\right)~.
  \label{Ecirc}
\end{align}
whereas the power emitted in gravitational waves is~\cite{Blanchet:2013haa}
\begin{align}
 \mathcal{P}_{\rm GW}^{\rm 3.5\,PN} &= \frac{32c^5}{5G_N}\nu^2 x^5 \biggl\{ 1 +
  \left(-\frac{1247}{336} - \frac{35}{12}\nu \right) x + 4\pi x^{3/2}
   + \left(-\frac{44711}{9072} +
  \frac{9271}{504}\nu + \frac{65}{18} \nu^2\right) x^2 +
  \left(-\frac{8191}{672}-\frac{583}{24}\nu\right)\pi x^{5/2}
  \nonumber \\ & \qquad \qquad \quad +
  \left[\frac{6643739519}{69854400}+
    \frac{16}{3}\pi^2-\frac{1712}{105}\gamma_\text{E} -
    \frac{856}{105} \ln (16\,x) 
     +  \left(-\frac{134543}{7776} +
    \frac{41}{48}\pi^2 \right)\nu - \frac{94403}{3024}\nu^2 -
    \frac{775}{324}\nu^3 \right] x^3 \nonumber \\ & \qquad \qquad
  \quad + \left(-\frac{16285}{504} + \frac{214745}{1728}\nu +
  \frac{193385}{3024}\nu^2\right)\pi x^{7/2} +
  O\left(\frac{1}{c^8}\right) \biggr\}~,
  \label{fluxEx}
\end{align}
where we used $\nu \equiv \mu/(M_1 +M_2) =  M_1M_2/(M_1 + M_2)^2$.
We can now solve the energy balance equation $dE_{\rm tot}^{\rm 3.5PN}/dt = - \mathcal{P}_{\rm GW}^{\rm 3.5PN}$, and compare the result with the deviation expected from the presence of a dark dipole radiation (see section~\ref{sec:DarkDipole}).
We show our result in fig.\,\ref{fig:PNDyn}. The inclusion of post-Newtonian corrections changes the prediction of gravity at the leading order in the opposite direction if compared to the electric dipole radiation (see fig.\,\ref{fig:Chirping}, right panel). This indicates that a more accurate analysis that includes also post-Newtonian corrections will not drastically change the qualitative conclusions outlined in section~\ref{sec:GW}.








\end{document}